\documentclass{aa}  
\usepackage{graphicx}
\usepackage{txfonts}
\usepackage{natbib}

\def\widerul{\vrule height 2.5ex width 0ex depth 0ex}

\def\asec{\ifmmode ^{\prime\prime}\else$^{\prime\prime}$\fi}

\def\it{\sl}
\def\degs{\ifmmode ^{\circ}\else$^{\circ}$\fi}
\def\amin{\ifmmode ^{\prime}\else$^{\prime}$\fi}
\def\asec{\ifmmode ^{\prime\prime}\else$^{\prime\prime}$\fi}
\def\fm{\hbox{$.\!\!^{\rm m}$}}            % Fractions of magnitudes
\def\farcs{\hbox{$.\!\!^{\prime\prime}$}}  % Fractions of arcseconds
\def\psr{PSR~J1124-5916}
\def\degs{\ifmmode ^{\circ}\else$^{\circ}$\fi}
\def\amin{\ifmmode ^{\prime}\else$^{\prime}$\fi}

\def\eqalign#1{\null\,\vcenter{\openup1\jot \m@th
   \ialign{\strut\hfil$\displaystyle{##}$&$\displaystyle{{}##}$\hfil
   \crcr#1\crcr}}\,}
\begin{document}
   \title{A likely optical counterpart      
   of the G292.0+1.8  pulsar wind nebula.
\thanks{Based 
%vk on observations collected at the European Southern Observatory, Paranal, Chile
%vk (ESO Programme 71.D-0499(A)).}}
on observations made with ESO telescope at the Paranal Observatory under
Programme 78.D-0617(A) and with archival ESO VLT data, obtained from
the ESO/ST-ECF Science Archive Facility.}}

%   \subtitle{I. Overviewing the $\kappa$-mechanism}
\author{S.V.~Zharikov\inst{1}
\and Yu.A.~Shibanov\inst{2} 
\and D.A. Zyuzin\inst{2}
\and R.E.~Mennickent\inst{3}
\and V.N.~Komarova\inst{4}
}

\offprints{S. Zharikov, \\  \email{ zhar@astrosen.unam.mx}}
 
\institute{
Observatorio Astron\'{o}mico Nacional SPM, Instituto de Astronom\'{i}a, Universidad Nacional Aut\'{o}nomia de Mexico, Ensenada, BC, Mexico, 
zhar@astrosen.unam.mx
\and
Ioffe Physical Technical Institute, Politekhnicheskaya 26,
St. Petersburg, 194021, Russia \\
shib@astro.ioffe.ru
\and
Departamento de Fisica, Universidad de Concepcion, Casilla 160-C, Concepcion, Chile \\
rmennick@astro-udec.cl
\and 
Special Astrophysical Observatory Russian Academy of Science, Nizhnii Arkhyz, 
Russia \\
vkom@sao.ru
%\and Isaac Newton Institute of Chile, SAO Branch, Russia 
}

%   \date{Received September 15, 1996; accepted March 16, 1997}

% \abstract{}{}{}{}{} 
% 5 {} token are mandatory
 
  \abstract
   {G292.0+1.8 is the Cas A-like supernova remnant containing the young 
pulsar PSR J1124-5916, which powers a compact torus-like pulsar wind 
nebula with a jet visible in X-rays.}
   {We have performed  deep optical 
observations  of the pulsar field  %in  an attempt  
to detect the optical counterpart of the pulsar and its %wind 
nebula. }
   {The observations were carried out using the direct imaging mode of FORS2 
at the ESO VLT/UT1 telescope in the $V$, $R$, and  $I$ bands. We also analyzed   archival images 
obtained with  the Chandra/ACIS-I, ACIS-S, and HRC-S in X-rays.
% chival data we revised  the PSR J1124-5916 X-ray 
%counterpart position using a dozen of USNO stars visible in the Chandra 
%FOV. 
}
   { In all three optical bands 
   we detect  a  faint  elliptical nebulosity,  %extended elliptical object, 
   whose   brightness peak and center position are    
consistent at a sub-arcsecond level with the X-ray position of the pulsar. 
The field is densely packed with background stars, but after  subtraction 
of these stars the morphology of the object and the orientation 
of its major axis appear to be in a good agreement with the brightest inner part  of the pulsar 
 nebula  torus region seen almost edge on in X-rays. Within the nebulosity we do not resolve any point-like 
optical object  that could be identified with the pulsar and 
estimate its 
%that the pulsar 
contribution  to the observed nebulosity flux as %less than 
$\la$20\%. Extracting the X-ray spectrum from the physical region equivalent to 
the optical source position and extent  and combining that with the measured optical fluxes, 
we compile a tentative multi-wavelength spectrum of the 
inner part of the  nebula. Within uncertainties of the interstellar extinction towards 
G292.0+1.8 it is reminiscent of either the Crab or PSR B540-69 and J0205+6449 
pulsar wind nebula spectra.}  
{The position, morphology, and spectral properties of the detected nebulosity  
 suggest that it is  the likely optical counterpart of the pulsar plus its 
wind nebula system in G292.0+1.8.  Higher spatial resolution optical observations and the extension of 
the  broad-band spectrum of the proposed counterpart candidate towards the IR and UV are necessary to confirm its origin.}
%   {The 
%position, morphology and spectral properties of the detected object strongly suggests 
%that we have detected  the optical counterpart of the pulsar plus its 
%wind nebula system in G292.0+1.8.}
%{}


\authorrunning{S. Zharikov, Yu. Shibanov, D.A. Zyuzin et al.}
\titlerunning{ A likely optical  counterpart }
   \maketitle

%
%________________________________________________________________
%%%%%%%%%%%%%%%%%%%%%%%%%%%%%%%%
\section{Introduction}
%%%%%%%%%%%%%%%%%%%%%%%%%%%%%%
\label{sec1}
Optical emission from radio pulsars  contains  important  information on  
the poorly
%not yet clearly 
understood physical processes responsible for the multi-wavelength 
radiation of rotation-powered neutron stars (NSs).  
So far only a small part ($\le$1\%) of the $\sim$1700  known radio  pulsars has been detected 
in the optical range \citep{mig05}. The most luminous 
%pulsars
of them
 are relatively young  and 
energetic to produce  a compact torus-like pulsar wind nebulae (PWNe), 
which are typically visible in X-rays. All these  young optical pulsars 
are associated with  
supernova remnants (SNRs):  Crab,  PSR B0540-69 ({ SNR B0540-69.3} in the LMC), 
PSR B1509-58 (G320.4-1.2), and Vela. Only for the  first two 
of these pulsars the surrounding PWNe 
has also been
%were also 
identified  in the optical range \citep[e.g.,][]{ser04}.
The multi-wavelength 
%power-law 
data
%spectra
 of the Crab-pulsar and its PWN  show the power-law spectra with 
%only 
one spectral break between the optical and X-rays, while  
the spectral energy 
%data
distribution 
for PSR B0540-69  and its PWN suggest 
a double-knee break in the same range. Recent optical \citep{shib08} and mid-infrared \citep{sla08} detections  
of the torus-like PWN around PSR J0205+6449 in the Crab-like  SNR 3C~58  
perhaps also suggest the double-knee break. 
Additional optical identifications  can provide new insights  on this issue.     

The young  PSR J1124-5916  has been only recently  discovered in 
the radio  \citep{Camilo}  and X-rays \citep{Hughes1,Hughes2}.   
It is associated  with  the SNR G292.0+1.8 (MSH 11-54) which is  the %known as a 
third  oxygen-rich  SNR known in the Galaxy
after Cas A and Pupis A. G292.0+1.8 has a morphology similar to Cas A 
on small scales \citep{hwan00}; however, no Fe K line emission was detected in its spectrum.  
As  the  Crab-pulsar,  PSR J1124-5916  powers a torus-like  X-ray PWN with a jet \citep{Hughes1,safi02,Hughes2,park07}. 
The pulsar period, 135 ms,  and the period derivative, 7.4$\times10^{-13}$, imply a  characteristic age of $\tau$$\approx$2900 yr, 
consistent with   2700--3700 yr age of the SNR \citep{Camilo,chev05}.  The derived spin-down luminosity of this pulsar is 
$\dot{E}$=1.2$\times10^{37}$ ergs~s$^{-1}$.  This ranks PSR J1124-5916 as the sixth youngest and 
the eighth most energetic  among the all rotation-powered pulsars  known. 
Because of its age, optical properties, and 
% the presence of the pulsar with the PWN, 
association with the pulsar and  PWN, % to the pulsar, 
G292.0+1.8 most closely resembles the oxygen-rich { SNR 0540-69.3} in the LMC. At the same time,  
the   spin-down energy loss of PSR J1124-5916  makes this pulsar 
closer  to the older Vela-pulsar which is known to be very under-luminous 
in the optical and X-rays.

The discovery of PSR J1124-5916 and its PWN in X-rays has 
prompted deeper  narrow-band and spectral optical studies of G292.0+1.8 \citep{wink06} 
%and
as well as
 broad-band searches %ing 
 for the pulsar optical counterpart  \citep{Hughes3}.  
Nevertheless, no counterpart has been detected with the CTIO 4 m telescope 
down to  { an estimated} %a moderate
% Serg N1
 stellar visual %upper 
 limit of $\sim$26.4 mag.
%Accounting for a 
Assuming distance of $d$$\approx$6~kpc  and interstellar reddening toward 
the pulsar of  $A_V$$\approx$1.8-2.2 mag., this implies   that the intrinsic optical 
luminosity  % limit $L_{\mathrm{opt}}$ for 
for PSR J1124-5916 must %has to 
be  lower than that %or
of
 the younger  
Crab, B0540-69, and B1509-58 pulsars, but  still  above of the older Vela-pulsar. 
%Ordering these young optical pulsars by  characteristic age and/or spindown luminosity  
%shows (Table~\ref{T.1}) that  
%The  optical luminosity  of J1124-5916 is likely to be between 
%the B1509-58 and the Vela-pulsar. 
This is also supported by the comparison of the X-ray luminosity %$L_{\mathrm{X}}$
%for
of this pulsar with  an empirical relation  between the optical and X-ray luminosities 
of the pulsars detected in both spectral domains 
\citep{zhar02,zhar04,zhar06,zav04}, suggesting further optical 
studies of the PSR J1124-5916 field.

Here we report  %on 
{ new} % Serg N2
 deep optical imaging of  {PSR J1124-5916} 
field in the $V$, $R$ and $I$ bands with the ESO Very Large Telescope (VLT).
The observations allowed us to find a likely candidate to the optical counterpart 
of the J1124-5916 pulsar+PWN system. We compare our optical images with the X-ray  data retrieved 
from the Chandra archive\footnote{ACIS-I, Obs 6677-6680, 8221, 8447, 530 ks exposure, PI S. Park}. 
The observations and  data reduction are described  in
Sect.~\ref{sec2}.  % Astrometric and photometric referencing   
%are given  in Sect.~\ref{sec3}. 
We present our results  in Sect.~\ref{sec3} and 
discuss them  in  Sect.~\ref{sec4}. 
%%%%%%%%%%%%%%%%%%%%%%%%%%%%%%%%%%%%%%%%%%%%%___________
\section{Observations and data reduction}  
%%%%%%%%%%%%%%%%%%%%%%%%%%%%%%%%
%Archive Data}
\label{sec2}
%%%%%%%%%%%%%%%%%%%%%%%%%%%
\subsection{Observations}
%%%%%%%%%%%%%%%%%%%%%%%%%%%%
The images of the pulsar field were obtained in the Bessel $V$, $R$ and $I$ bands 
with the FOcal Reducer and low dispersion Spectrograph (FORS2\footnote{For the instrument details 
see {www.eso.org/instruments/fors/}}) at  the VLT/UT1 (ANTU) unit  during several service mode  runs 
in  January and February 2007. A standard resolution mode was used with an image scale of  
 $\sim$0\farcs251/$\mathrm{pixel}$  and a field of view (FOV) of $\sim$7\amin$\times$7\amin. 
{ Because G292.0+1.8 and its pulsar are located near the Galactic plane, %region, 
which is densely packed with stars, we used  %Therefore, } 
a FORS2 occulting-bar setup for our observations.  This allowed us   %was used
 to minimize  contamination    
over the pulsar region  by illumination and saturation spikes from bright field stars 
surrounding the pulsar (see Fig.\ref{fig:1}, {\sl left panel}).     }
Sets of 3--10 min. dithered exposures were obtained in each of the bands with the total integration time 
of 8925~s, 7140~s, and 2200~s  in  the $V$, $R$ and $I$ bands, respectively. 
The observing conditions were rather stable with  seeing  values varying from 0\farcs6 to 0\farcs85.
The {\sl Log} of the observations is given in Table~\ref{t:log}. 
Standard data reduction including  bias subtraction, flat-fielding, and cosmic ray traces removing  
was performed making use of the {\tt IRAF} and {\tt MIDAS}  tools. 
%%%%%%%%%%%%%%%%%%%%%%%%%%%% Table 1 %%%%%%%%%%%%%
\begin{table}[t]
\caption{Log of the VLT/FORS2 observations of 
{\psr}. }
\begin{tabular}{llcll}
\hline\hline
 Date              &  Band, dithered  & Exposure       &  Airmass         & Seeing        \\
                   &  Exposure number & length, $s$          &                  &$arcsec.$   \\
   \hline \hline                
    2007-01-26     &  $V$, 10       &  595       &  1.22     &    0.6              \\
    2007-01-27     &  $V$, 5        & 595        &   1.22      &  0.6          \\
                   &  $I$, 7        & 200        &     1.24    &  0.6        \\                               
2007-02-20         &  $R$, 12       & 595        &   1.24       &  0.8               \\
                   &  $I$, 4       & 200        &    1.29     &     0.85         \\
\hline
\end{tabular}
\label{t:log}
\end{table}
%%%%%%%%%%%%%%%%%%%%%%%%%%%%%%end Table 1 %%%%%%%%%%%%%
%%%%%%%%%%%%%%%%%%%%%%%%%%%%%%%%%%%%%%%%%%%%%
\subsection{Astrometric referencing}% of optical data} 
\begin{figure*}[t]
 \setlength{\unitlength}{1mm}
 \resizebox{12.cm}{!}{
 \begin{picture}(120,85)(0,0)
\put (90,0) {\includegraphics[width=92mm, bb=120 240 440 530, clip=]{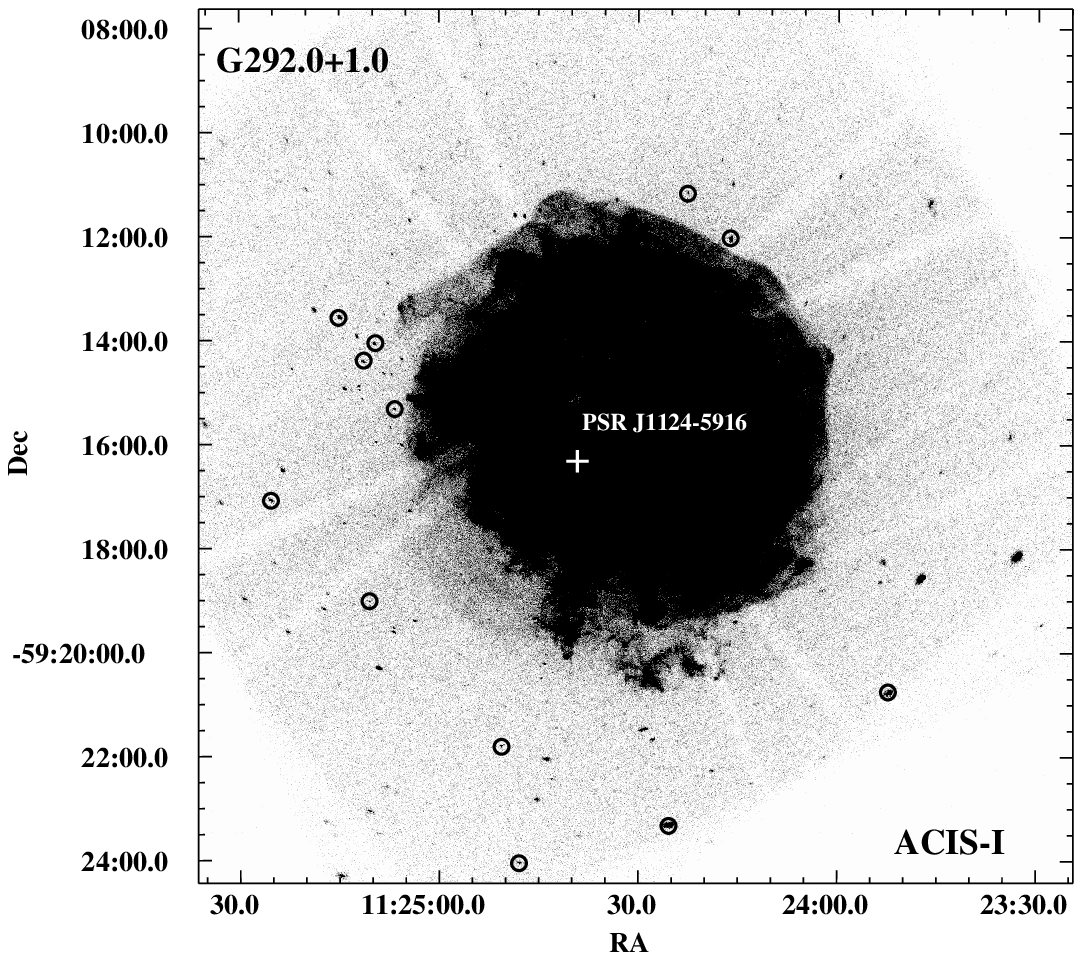}} 
\put (0,0) {\includegraphics[width=90mm,bb=129 240 460 550, clip=]{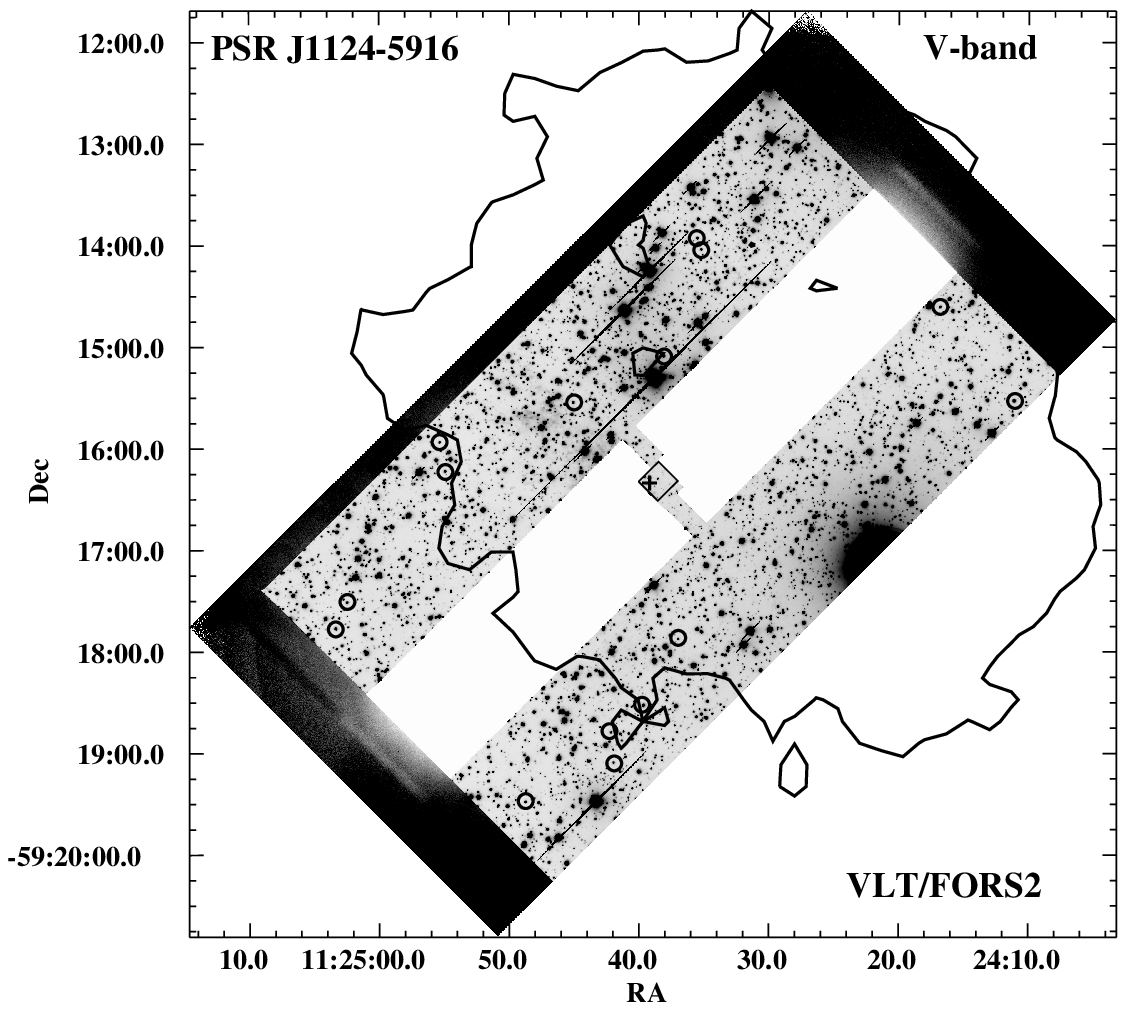}} 
 \end{picture}}
 \caption{The optical VLT/V-band  ({\sl left}) and the X-ray Chandra/ACIS-I ({\sl right}) images of the  PSR J1124-5916/SNR G292.0+1.8   
 field. White rectangular regions on the VLT CCD chip are covered by  
 occulting bars to eliminate the contamination in the pulsar neighborhood by the light from bright field stars.
%  within a small  box 
 %in the chip  center by the light from bright field stars. 
  An outermost SNR contour from the X-ray image is overlaid 
 on the optical image.  Circles  mark the USNO stars used for astrometric referencing of the images, the
 cross marks the pulsar position. 
 The small box in the center of VLT/V-band image  is enlarged in Fig.~\ref{fig:2}.}
 \label{fig:1}
 \end{figure*}

%%%%%%%%%%%%%%%%%%%%%%%%%%%%%%%%%%%%%%%%%%%%%
 Astrometric referencing was applied to the summed VLT images in  each of the bands.  
To obtain a precise astrometric  solution,  the positions of the astrometric standards selected from the USNO-B1 
astrometric catalog\footnote{USNO-B1 is currently
incorporated into the Naval Observatory Merged Astrometric Data-set
(NOMAD) which combines astrometric and
photometric information of Hipparcos, Tycho-2, UCAC, Yellow-Blue6, USNO-B,
and the 2MASS, www.nofs.navy.mil/data/fchpix/}
were used as  reference. 
Thousands of  USNO-B1 reference objects can be identified in our FOV, 
which is extremely crowded by stellar objects.  
Recent release of the Guide Star Catalog 
(GSC-II v2.3.2)\footnote{www-gsss.stsci.edu/Catalogs/GSC/GSC2/GSC2.htm}
provides practically the same number of the standards  but contains 
no information on proper motions and the declared astrometric errors (0\farcs3) are
higher than the nominal 0\farcs2 uncertainty of the USNO-B1.
A number of stars from the UCAC2 catalog is also present 
but they are saturated in our images. We discarded the reference stars with significant 
proper motions and catalog positional uncertainties $\ga 0\farcs4$ along with the saturated ones. 
Finally, to minimize  potential positional uncertainties 
caused by overlapping stellar profiles in the crowded FOV, we selected only 15 isolated stars 
(see Table~\ref{t:astr} and  Fig.~\ref{fig:1}, { \sl left panel}). Their pixel coordinates %of these  objects
%, considered to be suitable astrometric reference points , 
were derived making use of the IRAF task {\it imcenter}. 
%Finally, 
The IRAF tasks {\sl ccmap/cctran} were 
applied for the astrometric transformation of the images.  Formal {\sl rms} uncertainties of the  astrometric
fit for our images are $\Delta$RA$\la$0\farcs07 and $\Delta$Dec$\la$0\farcs05  for each 
of the bands, and  the fit residuals are $\la$0\farcs13, which is compatibile with the maximal catalog position 
uncertainty of the selected  standards. 
Selection of another set of isolated reference objects does not change significantly 
the result and a conservative estimate of our 1$\sigma$ astrometric referencing uncertainty is 
$\la$0\farcs15 in both RA and Dec for all three optical bands. 
  
For further comparison of the optical and  X-ray images  
we have also performed  astrometric referencing of the  G292.0+1.8 Chandra/ACIS-I archival 
X-ray images. We combined the multiple data sets obtained  with the ACIS-I  making use of the {\it merge\_all v3.6} 
script of the CIAO tool. The resulting image  %is deeper than the individual OBs,   that 
allowed us to resolve tens point-like objects located outside the SNR X-ray boundary (Fig.~\ref{fig:1}, {\sl right panel}). 
Most of them can be identified with the USNO standards. Potential reference objects within the SNR were ignored 
to  exclude  any false identification with numerous  X-ray  knots and filaments within the SN remnant. 
%This increases the reliability of the referencing as compared to  the case 
%when the reference objects are within the SNR and their false identifications   
%with numerious  X-ray  knots and filaments within the remnant cannot be excluded.  
Using similar selection criteria 
as in the optical case, we finally selected  12 suitable standards,  listed in Table~\ref{t:astr} 
and marked in Fig.~\ref{fig:1} ({\sl right panel}). The resulting  {\sl rms} uncertainties of the respective  
astrometric fit %, also performed with the IRAF {\sl ccmap/cctran}, 
are $\Delta$RA$\approx$0\farcs187 and 
$\Delta$Dec$\approx$0\farcs122 with the fit residuals of $\la$0\farcs27. This is compatible with the maximal 
positional uncertainty of the selected standards and makes the quality of the astrometric referencing of the ACIS-I  image   
comparable to that for the VLT images. The resulting shift between the initial and transformed X-ray images was   
$\approx$0.4 of the ACIS pixel size (0\farcs5) or about 0\farcs2 mainly directed towards the east. This ensures 
%us in 
an almost perfect pointing accuracy of the X-ray observations of the G292.0+1.8 field with the ACIS-I.         
%%%%%%%%%%%%%%%%%%%%%%%%%% Fig 1 %%%%%%%%%%%%%%%%%%%%%%%%%%%%
%%%%%%%%%%%%%%%%%   end Fig 1 %%%%%%%%
%%%%%%%%%%%% Table 2 %%%%%%%%%%%%%%%%
\begin{table}[t]
\caption{USNO-B1.0 stars used for astrometrical referencing with coordinates 
(epoch J2000.0) and their 1$\sigma$ errors.}
\begin{tabular}{lcccc}
\hline\hline
 Name &  RA  & Dec &$\sigma_{RA}$ & $\sigma_{Dec}$ \\
      & hh mm ss &     dd mm ss& mas & mas \\ \hline
  &     VLT                        &    images                      &      &     \\
 0307-0263383 &11 24 35.5253 &-59 13 55.840 & 18 & 23 \\      
 0307-0262730 &11 24 16.8207 &-59 14 36.360  &92 & 79  \\    
 0307-0262544 &11 24 11.1133 &-59 15 31.800 & 10 & 39 \\
  0306-0262602 &11 24 48.7647 &-59 19 28.770 & 75 & 23  \\
  0307-0263370 &11 24 35.2033 &-59 14 02.500 & 44 &163 \\
 0307-0263484 &11 24 38.1093 &-59 15 05.450  &44 & 54 \\
 0307-0263768 &11 24 45.0667 &-59 15 32.820  &92 &228   \\
 0307-0263438 &11 24 36.9907 &-59 17 52.570  & 9  &77  \\
0306-0262194 & 11 24 39.6380 & -59 18 31.850  & 63 & 40 \\
 0306-0262311 & 11 24 42.1680 & -59 18 47.830 & 148 & 36 \\
 0306-0262297 & 11 24 41.9627 & -59 19 05.730 & 56 &106 \\
 0307-0264210 &11 24 54.9427 &-59 16 13.580  &40  &37 \\
 0307-0264219 &11 24 55.3613 &-59 15 56.250 &138 & 40   \\ 
 0307-0264485 &11 25 02.5093 &-59 17 30.950 &122 &100 \\ 
0307-0264529 &11 25 03.3600 &-59 17 46.450  &50  & 79 \\  \hline
        &  Chandra/ACIS-I                           &  image                          &       &       \\
 0307-0294705         &       11 25 06.7700  &  -59:15:19.710  &  171  &	74   \\
  0307-0292657        &	 11 24 16.0510  & -59:12:02.614   &  12    &  12   \\  % 8629-00683-1 11 24 16.0380 -59 12 02.590   0   0
 0306-0287673         &      11 24 25.3659  & -59:23:19.419   &  43     & 56    \\   % 11 24 25.3960 -59 23 19.200  37  24
0306-0286344          &    11 23 52.2471  & -59:20:46.686   &   15   &   15   \\   % 0306-0260358 11 23 52.0220 -59 20 46.790 999 489
 0308-0292285  	&    11 24 22.6380  & -59:11:09.990   &   32   &   59   \\   %0308-0263739 11 24 22.6380 -59 11 09.990  32  59 
 0307-0294814 	&    11 25 09.6299  & -59 14 04.161   &   0    &   0	\\	   
 0307-0295391         &    11 25 25.2547  & -59 17 04.560   &  200 &  200    \\  
0305-0283993          &    11 24 47.9537  & -59 24 02.740   &  0     &  0	\\	   
0306-0288792  	&    11 24 50.6720  & -59 21 48.220   &   122  &   50	 \\	% 0306-0262690 11 24 50.6720 -59 21 48.220 122  45   
 0306-0289663 	&    11 25 10.4193  & -59 19 01.500   &   50    &  184   \\  % 0306-0263528 11 25 10.4193 -59 19 01.500  47 567 
 0307-0294870 	&    11 25 11.3865  & -59 14 24.310   &   0     &  0	\\	   
 0307-0295000 	&    11 25 14.9267  & -59 13 33.540   &   313  &   61	\\	   
\hline
\end{tabular}
\label{t:astr}
\end{table}
%%%%%%%%%%%%%%%%   end Table 2 %%%%%%%%%%%%
%%%%%%%%%%%%%%%%%%%%%%%%%%%%%%%%
\subsection{Photometric calibration.}                
%%%%%%%%%%%%%%%%%%%%%%%%%%%%%%%%%%%%%%%%%%%%%%%%%%%%%%
The observing conditions during our observations were photometric. 
The photometric calibration was carried out  with {PG1323-086}, {PG1633+099},
{PG0231+051},  and {RUBIN149} photometric standards from \citet{Landolt}  
observed during the same nights as our target. We fixed the atmospheric extinction coefficients at  their mean values  
adopted from  the VLT home page\footnote{ http://www.eso.org}: k$_V$=0\fm11$\pm$0.01, k$_R$=0\fm03$\pm$0.01, 
and k$_I$=0\fm00$\pm$0.01. The first observing run in 2007-01-26, when the standards were observed at significantly 
different airmasses, allowed us to check that the atmospheric extinction during our observations was really in  agreement 
with the mean one. The photometric fits performed for each of the observing nights  
showed that the photometric zero-points  have not varied significantly from night to night 
despite of a day and even a month gaps in the integration times for the $V$ and $I$ bands, 
respectively (Table~\ref{t:log}). As a result,  we obtained the following  magnitude zero-points for the summed 
images $V^{ZP}$=27\fm83$\pm$0.02, $R^{ZP}$=27\fm95$\pm$0.02, and $I^{ZP}$=27\fm31$\pm$0.02, 
where the errors account for the statistical uncertainties of magnitude measurements, 
the extinction coefficient uncertainties, 
and marginal zero-point variations from night to night. 
%, if the data for the target in a particular band were collected  in different nights. 
The formal  $3\sigma$  detection limits of a point-like 
object in the co-added images for an one-arcsecond aperture are $V\approx 27.8$,  $R\approx 26.9$, and $I\approx25.7$.  
%Resulting stellar magnitudes can be transformed into  fluxes in physical units 
%using zero-points provided by \citet{Fukugita}.  
%
%$M_j$  can be transformed into  absolute fluxes $F_j$ 
%(in erg~s$^{-1}$~cm$^{-2}$~Hz$^{-1}$), whenever it is necessary,  
%using standard equations  $\log F_j = - 0.4 \left(M_j + M_j^0\right)$
%with the zero-points provided by 
%(\citet{Fukugita}): $M_V^0=48.613,\ M_R^0 = 48.800$, and $M_I^0=49.058$.
%%%%%%%%%%%%%%%%%%%  Fig-2   %%%%%%%%%%%%
    \begin{figure*}[t]
 \setlength{\unitlength}{1mm}
% \resizebox{15.cm}{!}{
% \begin{picture}(150,150)(0,0)
%\put (0,0) {
\includegraphics[width=184mm, bb= 38 214 517 622,clip=]{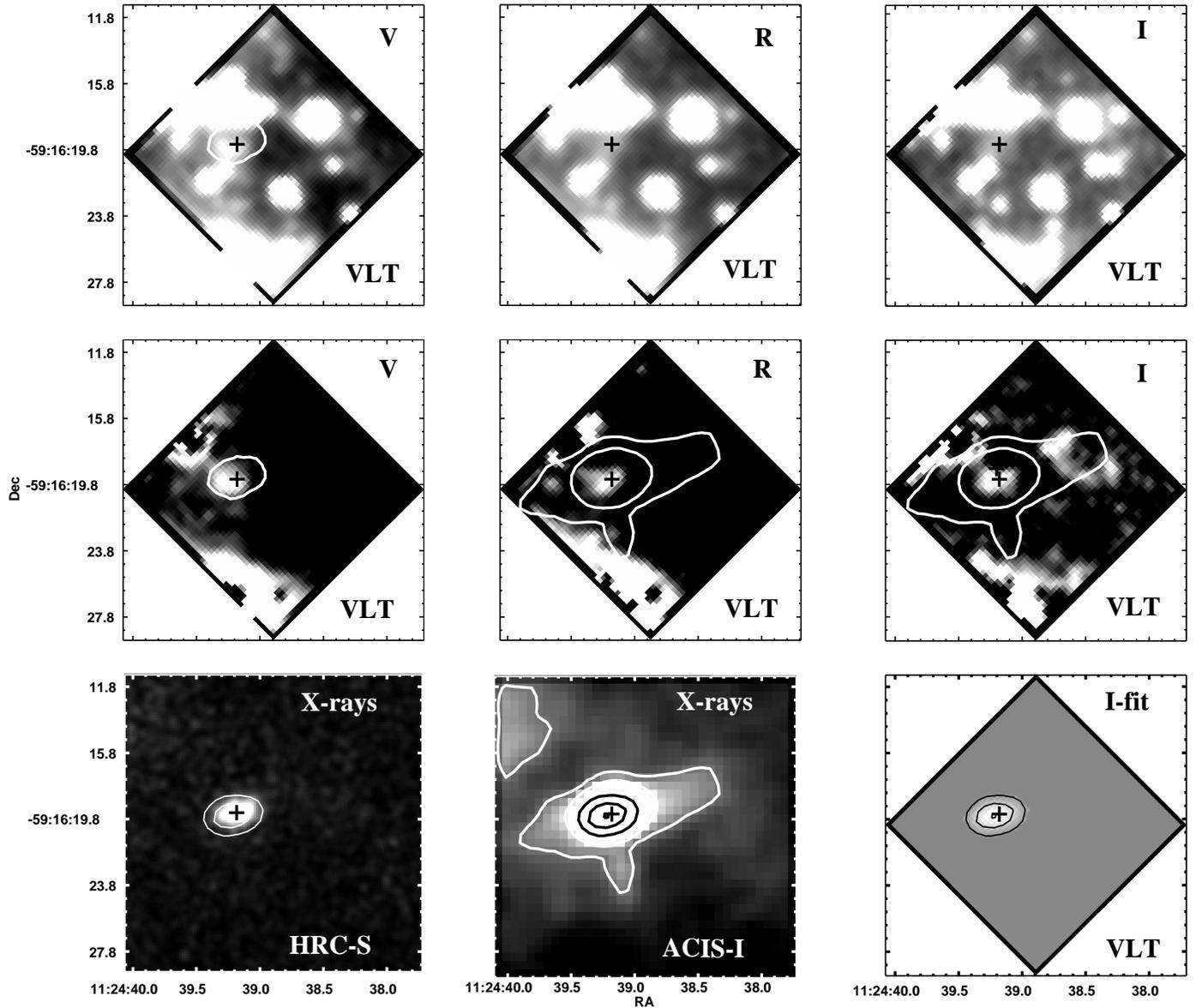}
%} 
% \put (106,50){\includegraphics[width=50mm, angle=-90, clip=]{Ibox.ps}} 
% \put (53,50){\includegraphics[width=50mm, angle=-90, clip=]{Rbox.ps}} 
 % \put (0,50){\includegraphics[width=50mm, angle=-90, clip=]{Vbox.ps}}
% \end{picture}}
 \caption{ Central  $\sim$16\asec$\times$16\asec~fragment of G292.0+1.8 containing the pulsar J1124-5916 and its PWN. 
 {\sl Top panels} are VLT/FORS2  $V$, $R$ and $I$  band optical images, as marked in the panels. 
 {\sl Middle panels} are the same images, where unsaturated stars have been subtracted. 
 {\sl Left and middle bottom panels} are the Chandra/HRC-S and ACIS-I X-ray images, respectively. 
   {\sl Right bottom panel} is the elliptical surface brightness fit of the optical counterpart candidate 
   to the pulsar+PWN system in the $I$  band. Its contours are overlaid on the X-ray images (black on ACIS-I and white on HRC-S).
   The contours outlining the boundary of the PWN in the HRC-S image and its brightness distribution in the ACIS-I image  
   are overlaid on the $V$ and the star subtracted  $R$ and $I$  images, respectively.  All images are smoothed with 
   a Gaussian kernel of three pixels. In the optical range this corresponds to a mean seeing value. Cross shows the X-ray position 
   of the pulsar and  sizes of its arms are 1$\sigma$  uncertainties of the position.   
       }
 \label{fig:2}
 \end{figure*}
%%%%%%%%%%%%%end Fig 2 %%%%%%%%%%%%%%
  %%%%%%%%%%%%%%%%%%%%%%% Fig 3  %%%%%%%%  
    \begin{figure}[t]
 \setlength{\unitlength}{1mm}
\includegraphics[width=84mm, bb= 38 174 507 582,clip=]{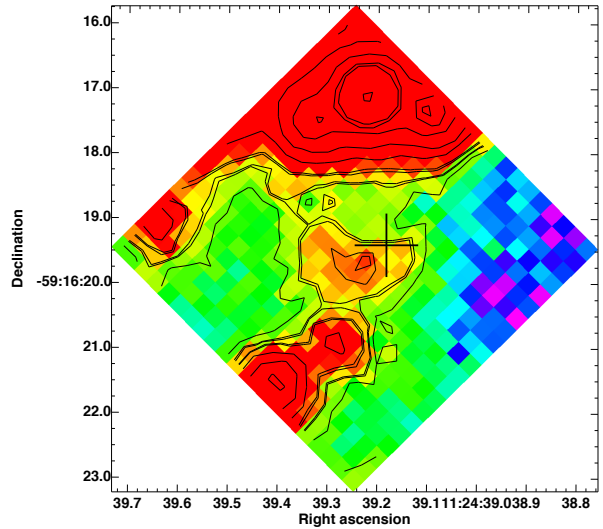}
 \caption{Zoomed fragment 
 %The   $\sim$7\asec$\times$7\asec~fragment 
 of the V band image shown in Fig.~\ref{fig:2}. %~containing the PSR J1124-5916 and its PWN with 
 Contours underline the surface brightness distribution.  The cross marks the X-ray position of  PSR J1124-5916,  
 the sizes of its arms are 2$\sigma$, or $\sim$0\farcs5, uncertainties of the position (see text for details).  
  %corresponds to $\sim 1^{\prime\prime} $ accuracy from Chandra data \citep{Hughes1} .
    }
 \label{fig:2a}
 \end{figure}
 %%%%%%%%%%%%%%%%%%%%%%%%%%%%%%%%%%%%% end Fig-3 %%%%%%%%%%
\section{Results}
\label{sec3}
%%%%%%%%%%%%%%%%%%%%%%%%%%%%%%%%%%%%%
\subsection{Detection of the pulsar/PWN  counterpart candidate.}
%%%%%%%%%%%%%%%%%%%%%%%%%%%%%%%%%%%%%5
The region containing the pulsar is enlarged in Fig.~\ref{fig:2}, where 
we compare our optical $V$, $R$, and $I$ images 
with the archival X-ray images obtained with 
%the
Chandra/ACIS-I and HRC-S\footnote{HRC-S, Obs 1953, 49.8 ks exposure, PI J. Hughes}. 
The dynamical range of the ACIS-I 
image was changed 
relative
%compared 
to that shown in  Fig.\ref{fig:1} to reveal the structure of the torus-like PWN with the south jet 
(assuming the torus is seen almost edge-on). Only the central part of the torus with the pulsar is seen on the less deep HRC-S image.  

{   In all three optical bands ({\sl top  panels} of Fig.~\ref{fig:2}), at the  pulsar  position,  marked by a cross, 
 we detect  a faint extended object, hereafter named  the nebulosity.  
To study its detailed 
 %For 
 spatial morphology 
 %details 
 we show  in Fig.~\ref{fig:2a} 
 %we show 
 a zoomed fragment 
 of the $V$ band image,  which is the deepest 
 available
 one and provides the best  spatial resolution   
% as compared to the other bands 
(cf., Table  \ref{t:log}).  % is shown in Fig.~\ref{fig:2a}. 
The spatial brightness distribution is underlined by contours, demonstrating  an %apparent 
east-west extent 
of  the nebulosity  and  a  distinct  brightness peak whose position   is compatible with the position 
of  the  pulsar in X-rays at $\sim$1$\sigma$ uncertainty level.  
  The north-south wings of the nebulosity  partially overlap with the wings  of  nearby north and south background stars, 
  specifically with the closest one  located  only  $\sim$1\farcs2 to the south of the pulsar. 
  However,   the  brightness of its central part  cannot be  caused %explained 
  by  
  the overlapping  wings of these stars.  The same  conclusion is valid for %also follows from 
  the  $R$ and $I$ images, 
  although they were obtained at about 30\% worse seeing and in these images the nebulosity   
  is  more  contaminated by the stellar wings. Nevertheless, its central peak is also clearly resolved. %there as well. }    
  }
  
  { Our limited spatial resolution and the presence  of the nearby stars does not allow us to make  definite conclusions 
  on the nebulosity extent on the original images.   To  better reveal the  morphology of the object   we  produced 
   star-subtracted images. 
   %To  better reveal the object morphology   we  subtracted the stars around the pulsar position.
    % produced the star subtracted images.   
    We constructed a model point spread function (PSF)  based  on a set of about 30 isolated 
  and unsaturated field  stars in our images and used the {\sl allstar} task from the IRAF {\sl digiphot} tool. 
  The task provides an efficient iterative algorithm for a careful subtraction  of  groups of stars with 
  overlapping PSFs. The latter  is crucial in our case since the most of the stars near the pulsar   
  are in  such groups.  Since the  subtraction process is  somewhat imperfect,   
  we  repeated it  several times varying the {\sl allstar} parameters. The results for each of the bands 
  were rather stable and  they are shown in  the {\sl  middle panels} of Fig.~\ref{fig:2}. The main contaminating factor,  
  the  star   $\sim$1\farcs2 south of the pulsar,  as well as other unsaturated 
  stars are perfectly subtracted.  That is obviously not possible for  saturated stars with corrupted PSFs,  
  whose remains are noticeable.   
  However,  they are  distant enough from the nebulosity  
   to affect  its morphology. 
%  As a result,  the nebulosity  shows   
  On the resulting images, the nebulosity has an ellipsoidal shape
  %an  ellipse-like shape  
  compatible  with the structure  seen in the HRC-S  image, 
  revealing only the brightest inner part of the PWN torus with the pulsar. 
  Both the extent of the optical source and the orientation of its major axis are similar 
  to what was found in X-rays. This suggests  that the detected source is a likely  optical 
  counterpart  of the central part of the  PSR J1124-5916$+$PWN system. }
  %,  with possible contribution from the pulsar itself.   
  %%%%%%%%%%%%%%%%% Table 3 %%%%%%%%%%%%%%% 
\begin{table}[t]
\caption{Parameters of the elliptical fit to the surface brightness of the 
PSR J1124-5916$+$PWN optical counterpart ordered by the major ellipse axis 
length (c.f., {\sl right bottom} panel of Fig.~\ref{fig:2}).}
\begin{tabular}{lllcll}
\hline\hline 
 Center$^a$ & Center$^a$  &  Major          & Ellip-      & PA$^c$       & $I$-band \\  %&  B mag$^c$  \\
  RA        & DEC         &   axis          & -ticity$^b$ &  W to N     & flux$^d$  \\  %&  in the   \\
                        &               &    length  &                    &                    &       \\                               
  (11:24:          & (-59:16:            &                &             &    &   \\  %& elliptical  \\  
  39.*)             &   19.*)   &  (arcsec)       &             & (degrees)  & (\%) \\  %&  aperture   \\
\hline                                                                                       
 216        & 60          &   1.26         & 0.20        & -1.3   & 24.9(7.6)   \\ % & 24.624(110)  \\
 230        & 59          &   1.50         & 0.25        & -0.2    & 46.4(8.1)  \\ % & 23.948(89)   \\
 240        & 61          &   1.76          & 0.37        &13.2   & 51.8(8.2)  \\% & 23.829(86)   \\
 235        & 59          &   2.00          & 0.34        &11.5   & 66.2(8.6)  \\ % & 23.563(79)  \\
% 218        & 59          &   2.259          & 0.525        &100.16   & 61.6(8.4)  \\% & 23.642(79)  \\
 211        & 57          &   2.50          & 0.50        &12.8   & 76.5(8.8)   \\% & 23.406(73)  \\
 212        & 57          &   2.76          & 0.48        &10.3   & 87.3(9.3)   \\% & 23.263(74)  \\
 212        & 57          &   3.01          & 0.48        &10.3   & 96.8(9.8)   \\% & 23.150(75)  \\
 212        & 57          &   3.21          & 0.48        &10.3   & 100(7.1)  \\% & 23.115(77)  \\                                       
\hline
\end{tabular}
\label{t:ell}
\begin{tabular}{ll}
$^a$~coordinates of centers of the ellipses (J2000);& \\
$^b$~defined as $1-l_{min}/l_{max}$, where $l_{min}$ and $l_{max}$ are the minor and  & \\
major ellipse axes lengths, respectively; & \\
$^c$~positional angle of the major axis;& \\ 
$^d$~flux from the elliptical aperture normalized to the flux from the  &  \\
largest aperture in this set. Numbers in brackets are 1$\sigma$ uncertainties.  &\\
\end{tabular}
 \end{table}
%----------------------------end of Table 3 --------------
%In the $V$ band the ellipticity of the optical object appears to be less obvious. 
%{ One of the reasons may be an imperfect  star subtraction distorting    
%outer regions of the source.}  
%
%The another one may be attributed to  a   
%strong  [OIII]  emission line from the remnant knots and filaments \citep{wink06} that enters into the $V$ passband.  
%Some  emission clamps visible in this band south of the suggested counterpart may be also associated with faint [OIII] knots. 
%Their presence   near the pulsar position cannot be excluded considering the available low spatial resolution  
%narrow-band [OIII] images published by \citet{wink06}. Another emission lines (e.g., H$_{\alpha}$, [OI], [OII] and [SII]) detected 
%from the remnant  in the red part of the spectrum enter into  the $R$ and $I$ passbands. However, they   
%are much fainter than that of [OIII]  and cannot significantly contaminate the respective parts of the continuum  emission  
%expected to be produced by the pulsar and its PWN. By these reasons, the  better coincidence of the shape  
%of the optical source in the red optical bands with the  X-ray PWN structure  is anticipated for 
%the real optical PWN counterpart. 

We do not resolve in our optical images any signature of the southern 
 X-ray jet of the PWN. This is not a surprise since the jet is  much fainter than the torus  and  it 
was clearly resolved in X-rays only  after a very deep observation 
with the ACIS-I \citep{park07}.

To compare quantitatively the morphology of the suggested optical counterpart  with the X-ray PWN torus structure, 
we have fitted the spatial intensity distribution in the star-subtracted images with a simple two-dimensional model 
(using the IRAF {\sl isophote} task) accounting for the brightness distribution by elliptical isophotes 
and taking into account the background level.  A similar fit has been made by  
\citet{Hughes2} to the pulsar/PWN source on the HRC-S image. The results of the  fit for the $I$ band are 
presented in  the {\sl right bottom} panel of Fig.~\ref{fig:2} and Table \ref{t:ell}. 
The fit shows that the brightest inner part of the nebula, within 1\farcs--1\farcs5 of the center, 
has almost a circular shape and emits  $\approx$20\%--40\%~ of the total flux. If our identification is correct, 
this can be considered as an upper limit of the pulsar contribution to the total pulsar+PWN flux. 
The coordinates of the  center defined from the fit are $\alpha_{2000}$=11:24:39.216, and $\delta_{2000}$=-59:16:19.60. 
The measured pulsar centroid positions in X-rays are $\alpha_{2000}$=11:24:39.183,  $\delta_{2000}$=-59:16:19.41 and $\alpha_{2000}$=11:24:39.180,  
$\delta_{2000}$=-59:16:19.49, as obtained  from the HRC-S\footnote{Our HRC-S coordinates, when rounded by the last two digit numbers, 
are compatible with the published ones \citep{Hughes2}} % : RA=11:24:39.10, Dec=-59:16:20.0
   and ACIS-I images, respectively. Both  are 
in a good agreement with each other and with the center position of the optical source  taking into
accounting the  $\sim$0\farcs15 and $\sim$0\farcs2 astrometric accuracy in the optical and 
X-rays  (Sect.~2.2).   The ellipticity of nebulosity  increases up to $\sim$0.5 for the outward optical nebula regions and the major axis is
slightly tilted from the RA axis  by about 10\degs--13\degs, which is also in  good agreement with about 13\degs~tilt 
seen in X-rays. In addition, the optical nebula is elongated almost symmetrically around its center with a total length of 
$\sim$3\farcs2, which agrees with the extent of the brightest part of the PWN seen in X-rays. 
Assuming that we see a tilted  circular torus, %using the elliptical fit 
we  estimate  
%that 
the torus  axis angle to the line of sight %is 
in a range of 50\degs--60\degs.
%using the elliptical fit.
%to explain the eccentricity of our fit.        
%%%%%%%%%%%%%%% Fig 3 %%%%%%%%%%
\begin{figure*}[t]
 \setlength{\unitlength}{1mm}
 \resizebox{12.cm}{!}{
 \begin{picture}(120,90)(0,0)
\put (0,0) {\includegraphics[width=87mm, bb=120 240 440 530, clip=]{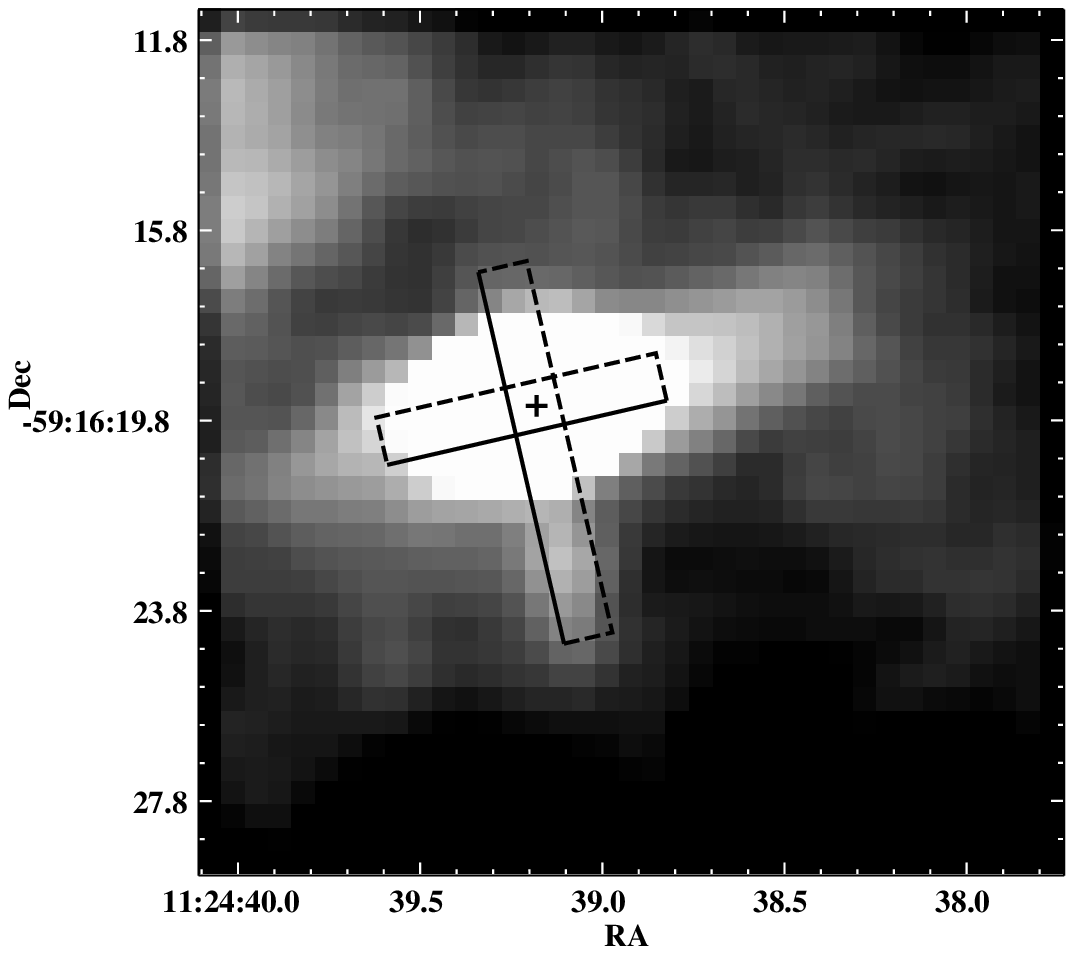}} 
\put (135,0) {\includegraphics[width=50mm, bb=30 77 335 574, clip=]{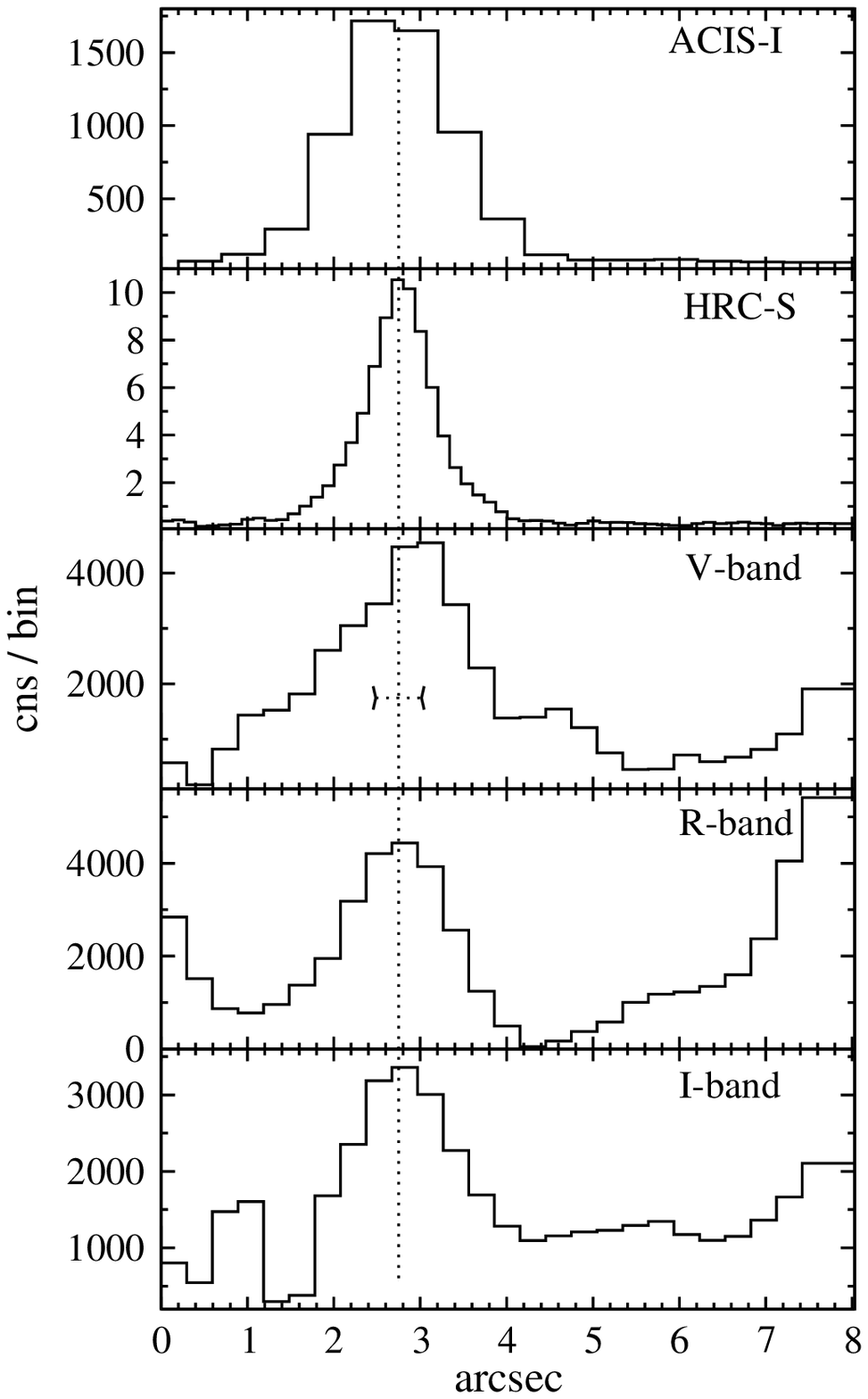}} 
\put (90,0) {\includegraphics[width=50mm,bb=30  77 335 574, clip=]{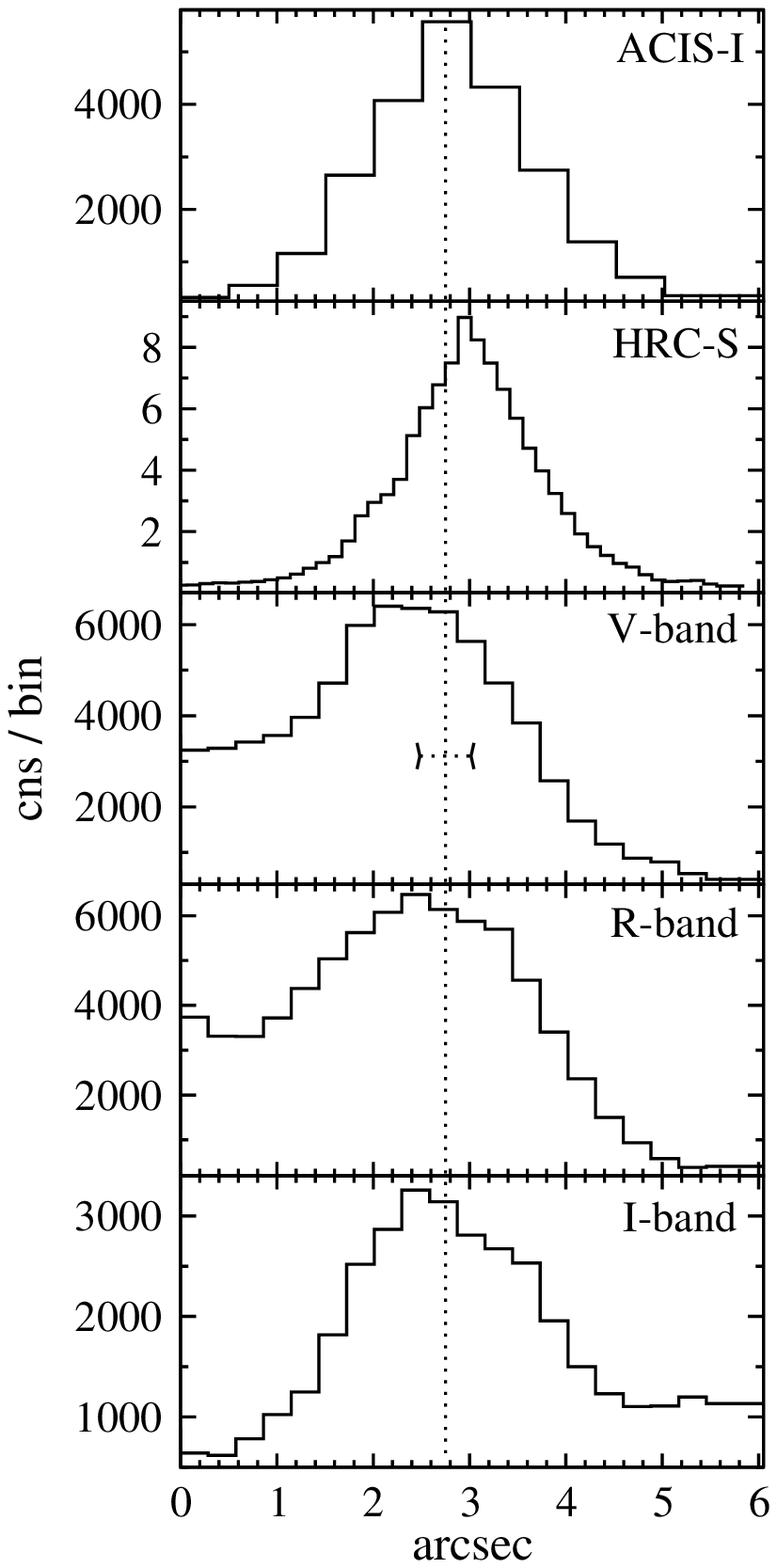}} 
 \end{picture}}
 \caption{ {\sl Left:} The ACIS-I image with positions of two slices 
 (bold lines) used for comparison of 1D X-ray  source profiles 
 of the J1124-5916 pulsar+PWN system 
 with its  optical counterpart candidate profiles; cross marks the position of the pulsar. 
 {\sl Middle} and  {\sl right}:  The source profiles extracted from the ACIS-I, HRC-S, $V$, $R$, and $I$ images, 
 as marked in the panels,  along the "horizontal"  and "vertical"  slices, respectively, shown in the {\sl left}. 
 The  coordinate origins of the horizontal axis in the {\sl middle} and {\sl right} 
 panels coincide with  the eastern and northern edges of the respective slices. Vertical dotted lines with horizontal error-bars mark 
 the X-ray position of the pulsar and its uncertainties.}
 \label{fig:3}
 \end{figure*}
%%%%%%%%%%%%%%%%%%%% end Fig 3 %%%%%%%%%%%%%%%

Extending the analysis of the nebular structure in the optical and X-rays we also considered 
 1D-spatial profiles of the images along the two slices directed along the nebula symmetry axes.  
The slice positions are shown in the {\sl left} panel of Fig.~\ref{fig:3}.   
Their widths are 1\asec, both centered on the the X-ray pulsar position.    
A "horizontal" slice  is 6\asec~long  and placed symmetrically over the pulsar position  
to include  most of the emission along the major axis of the nebula. Its PA to the RA axis is  $\approx$13\degs. 
The "vertical" slice is  orthogonal to the first one, it is 8\asec~long  
and extended  towards south-south-west to include also the X-ray jet. 
The respective 1D-profiles performed on the star-substracted images are shown in the {\sl middle} and {\sl right} panels of Fig.~\ref{fig:3}
%\footnote{ The optical %bands
%%the 
%profiles are extracted from the star subtracted images.}.
%The horizontal axis in the {\sl middle} panel is directed from the east to the west slice edges, 
%and in the {\sl right} panel from the north to the south. 
%Vertical dotted lines with horizontal error-bars 
%show the X-ray pulsar position and its uncertainties. 
These show the coincidence in the positions  of the main  peaks of the nebula and their extents 
in the optical and X-rays. 
The ACIS-I peak is broader than that of the HRC-S because of 
a lower spatial resolution of ACIS. 
 A wing asymmetry  of the $V$ and $R$ peaks  in the {\sl middle} panel 
and the flux enhancement in the southern part  of the all optical profiles  in the {\sl right} panel are not related 
to the nebula. These are  subtraction remains of nearby background stars east and south of the PWN, respectively. 
The southern jet, barely resolved  in the ACIS-I profile in the {\sl right} top panel, cannot be reliably resolved    
in the optical profiles in part, because of %partially owing to 
the asymmetry of the contaminating effect of the above remains.   
The FWHM of the optical PSF is $\sim$0\farcs6--0\farcs85 (Sect.~2.1), and the brightest
part of the nebula, which  extends to several arcseconds, is clearly resolved.
Although the position of the main optical peak coincides with the pulsar 
position, we do not resolve any point-like source at this position.  
To resolve the point-like pulsar from the nebula would require deeper imaging 
at higher spatial resolution, and/or time resolved observations.

Summarizing this part we can conclude that our spatial analysis %of the detected optical 
%nebula  show  that it is the very likely optical counterpart 
strongly favors the suggested optical identification of the G292.0+1.8 pulsar/PWN system. 
{ The main uncertainties  are related with a possible  imperfect subtraction of the  closest stars 
contaminating  the emission of the outer parts of the PWN optical counterpart candidate.   This may  affect  
 the derived extension of the optical  source along the minor axis of the X-ray PWN torus and 
 the parameters of the elliptical surface brightness fit.  However, the similarity  of the counterpart 
 candidate morphology in the star-subtracted images in the different  optical bands, where the  star appearances are considerably 
 different,  supports the correctness of the  performed subtraction  and  the PWN identification.
 Higher spatial resolution imaging is needed to confirm this. }   
%%%%%%%%%%%%%%%%%%%%%%%%%%%%%%%%%%%%%%%%%%%%%%%  
\subsection{Photometry}
%%%%%%%%%%%%%%%%%%%%%%%%%%%%%%%%%%%%%%%%%%%%
The optical photometry of the suggested counterpart %candidate 
was performed on the star-subtracted 
images. 
%({\sl middle } panels of Fig.~\ref{fig:2}). 
We used the elliptical apertures 
from the surface brightness fit described in Sect.~3.1. Their parameters %of the ellipses 
are presented in Table~\ref{t:ell}.   
The background levels were estimated from a circular annulus with   $\sim$7 pixel inner radius and a width of $\sim$5 pixels 
centered on the nebula center. The relative photometric errors were minimal ($S/N\approx 13$) for  apertures   
formally encapsulating $\ga$90\%~ of the total  nebula flux. The measured integral magnitudes of the  
presumed pulsar+torus nebula  are summarized in Table~\ref{t:phot}, %$V$=24\fm29$\pm$13, $R$=24\fm12$\pm$13, %$I$=23\fm12$\pm$13, 
where the errors include  the calibration zero-points uncertainties,  statistical measurement errors, and  
{  account for   
possible star subtraction uncertainties based on the magnitude 
dispersion after  the reiterations of the  subtraction process. } 
 We also tried circular and 
polygonal apertures  to better encapsulate the whole flux, but got practically the same results. 
The stellar magnitudes were transformed into fluxes  using zero-points provided by \citet{Fukugita}. 

{ The observed color indices of the nebulosity are $V-R=0.2(2)$ and $R-I=1.0(2)$.
We also measured  the indices of  a random set of faint field 
stars %-like objects  
located, as our nebulosity, outside the known emission line regions of the SN remnant  \citep{wink06,park07}. 
We obtained indices in the range of  $0.6 \le V-R\approx R-I \le 1.1$.  This  corresponds to the stellar "reddened"  color indices 
and it is significantly different from the  nebulosity colors.   
 %considerably different  
  %We found  that   
  %the nebulosity colors
 %are considerably different from the stellar ones, $0.6 \le V-R\approx R-I \le 1.1$, 
 %which correspond to the stellar "reddened"  color indices. 
   %\footnote{}
Based that, the detected nebulosity    
 can hardly be a result of a superposition of a few  overlapping,  unresolved faint background  
 stars. 
}
%The results are summarized in Table~\ref{t:phot}. 
%%%%%%%%%%%%  Table 4  %%%%%%%%%%%%%%%%555
\begin{table}[t]
\caption{Observed magnitudes and fluxes for the presumed optical  
pulsar/PWN counterpart in G292.0+1.8,  as well as de-reddened fluxes for different $A_V$ 
values.}
\begin{tabular}{llllll}
\hline\hline 
Band               & Mag.       &  log F          &                & log F &   \\ 
%($\lambda_{eff}$) 
 &  obs$^a$  &  obs$^a$          &     &  dered$^{a,b}$ &             \\  
\cline{4-6}
                   &            &                    & $A_V$=1.86  &  2.06(4) & 1.86--2.1           \\ 
%\cline{4-6}          
%              &            &                    &                       &                       &         \\ 
%($\mu$m)    
       & (mag)      & ($\mu$J)           &   &($\mu$J) &             \\ 
\hline 
%       &             &                   &            &           &   \\
% $V$          & 24.29(13)  & -0.161(52)         & 0.656(69)  &0.576(52)  & 0.656($^{+69}_{-132}$)             \\ 
%         &             &                   &            &           &                                    \\        
 $V$          & 24.29(13)  & -0.16(5)         & 0.58(5)  &0.66(7)  & 0.66($^{+7}_{-13}$)             \\ 
 %        &             &                   &            &           &                                    \\     
% $R$          & 24.12(13)  & -0.168(52)         & 0.507(68)  & 0.442(52) &  0.507($^{+68}_{-207}$)         \\ 
%(655)         &             &                   &            &           &                                    \\  
 $R$          & 24.12(13)  & -0.17(5)         &  0.44(5) & 0.51(7) &  0.51($^{+7}_{-22}$)         \\ 
  %       &             &                   &            &           &                                    \\  
% $I$          & 23.12(13)  & 0.129(52)          & 0.661(63)  & 0.609(52) &  0.661($^{+63}_{-104}$)           \\ 
   %      &             &                   &            &           &                                    \\ 
 $I$          & 23.12(13)  & 0.13(5)          & 0.61(5)  & 0.66(6) &  0.66($^{+6}_{-10}$)           \\ 
    %     &             &                   &            &           &                                    \\ 
\hline
\end{tabular}
\label{t:phot}
\begin{tabular}{ll}
$^a$~numbers in brackets are 1$\sigma$ uncertainties referring to last & \\
significant  digits quoted. & \\ 
$^b$~uncertainties include  $A_V$ uncertainties.      & \\
\end{tabular}
\end{table}
%%%%%%%%%%%%%%%%%%%%% of Table 4 -------------- 
%%%%%
%%%%%
%%%%%
%%%%%%%%%%%%%%%%%%%%%%
\subsection{Multi-wavelength spectrum of the pulsar+PWN source}
%%%%%%%%%%%%%%%%%%%%%%%%%%%%%%%%%%%%%%%%55
Using the optical fluxes for the pulsar/PWN counterpart candidate together  
with the X-ray data we can construct a tentative multi-wavelength 
spectrum of the central part of the source. 

To compare  our optical fluxes 
with the X-ray spectrum from the same physical region of the pulsar+PWN system 
we applied the 3\farcs2 major axis length elliptic aperture from Table~\ref{t:ell} 
that  encircles   $\sim$100\%~ of the optical flux.  
Using the ACIS-I Obs Ids 6677 and 6679  with the longest individual exposures 
($\sim$161 and $\sim$156 ks) of the G292.0+1.8 center and 
the CIAO {\sl acisspec} tool, 
within this elliptical aperture  we obtained  
7659 and 7393 source counts, respectively.  A circular region of $\sim$10\asec~diameter 
located  $\sim$16\asec~south of the pulsar was used to estimate the background  
 contribution, which was  $\la$1\% within the source aperture. 
Spectral data of both OBs were grouped to provide a minimum 25 counts per 
 spectral bin and fitted simultaneously by an absorbed power-law model using standard XSPEC tools. 
As a result, we obtained a statistically acceptable fit with a spectral index $\Gamma$=1.851$\pm$0.027,
absorbing column density $N_{H}$=(0.369$\pm$0.014$)\times$10$^{22}$~cm$^{-2}$ and normalization 
constant C =(1.38$\pm$0.08)$\times$10$^{-4}$~photons~cm$^{-2}$~s$^{-1}$~keV$^{-1}$. 
The fit had $\chi^2=0.962$ per degree of freedom and the unabsorbed integral flux  
was 5.626$\times10^{-13}$~erg~cm$^{-2}$~s$^{-1}$ in 0.3--10 keV  range. 

The fit residuals  do not show any signs of emission lines, as found by 
\citet{Hughes1}  with a few arcseconds larger aperture. This means that the spectrum 
%obtained 
is unlikely to be contaminated by the emission form the supernova ejecta. 
The derived spectral index is also compatible 
with typically observed indices for  Crab-like PWNe. We checked our result using the same aperture 
and  counts  per bin setup on previous  $\sim$44 ks observations of G292.0+1.8 with 
the ACIS-S\footnote{OBs ID 126, exposure 44 ks, PI G. Garmaire}. Within larger uncertainties resulted from   
lower count statistics at  this short exposure we obtained the same fit parameters. 
This ensured us that for the selected pulsar+PWN region  both the ACIS-I and the ACIS-S 
provide the same results.  
Finally,  we adopted a circular aperture of  2.5~pixel radius (1\farcs23) centered on the pulsar position, 
as used by \citet{Hughes1}, to estimate the spectrum of the pulsar on the ACIS-S data. 
Within uncertainties, both the ACIS-S and the ACIS-I  gave the  spectral index and 
the column density which are consistent with those obtained for the  elliptical aperture. 
For instance, for the ACIS-I we got $\Gamma$=1.851$\pm$0.027 and 
$N_{H}$=(0.367$\pm$0.014$)\times$10$^{22}$~cm$^{-2}$. 
This is in contrast to smaller values, $\Gamma$=1.72$\pm$0.05 and  
$N_{H}$=(0.317$\pm$0.015)$\times$10$^{22}$~cm$^{-2}$, published by \citet{Hughes1}. 
We have been able to reproduce the published values  only for the case 
of the ACIS-S data when the spectral range at the fit was constrained by 0.3--7 keV. A similar constraint 
for the ACIS-I data with much higher count statistics does not lead to any significant spectral hardening 
and/or $N_{H}$ decrease.  This shows that the PWN contribution to the total pulsar+PWN flux is still rather 
high even for such a small aperture and the previous estimates of the pulsar spectrum and $N_{H}$  based  on the short  
ACIS-S exposure may be %not quite 
unreliable.

The next step to the multi-wavelength spectrum  analysis  is the correction of the optical fluxes 
for  interstellar reddening. 
Compiling  various  hydrogen column density measurements in the radio and X-rays and  
comparing  the observed optical spectral-line intensity ratios for G292.0+1.8 filaments 
with theoretical models of the post shock line emission from the material ejected from 
the SN progenitor, \citet{wink06} adopted the interstellar color excess 
of $E(B-V)$=0.6 mag. On the other hand, using the $N_H$ value derived from our  
X-ray spectral fit and an empirical relation between  the effective  $N_H$  of the X-ray absorbing gas 
and the dust extinction $N_{\rm H}/E(B-V) = (5.55\pm0.093)\times 10^{21}$~cm$^{-2}$~mag$^{-1}$ \citep{pred95}, 
we obtain a larger value of $E(B-V)$=0.665$\pm$0.014 mag, which corresponds to the $V$-band reddening correction 
$A_V$=2.062$\pm$0.043 mag, adopting a standard ratio $A_V/E(B-V)=3.1$.  
Within 5$\sigma$ this is consistent with the above 
$E(B-V)$ value ($A_V$=1.86 mag) and  significantly smaller than the 
extinction of 0.86 mag caused by whole Galaxy 
%entire Galactic excess of 
 in  this
direction % provided by 
\citep{schleg98}. % for the direction towards G292.0+1.8.   
De-reddened optical fluxes for the considered  $A_V$  values and their combination  
are presented in Table~\ref{t:phot}. 
%(The choice of $A_V$=3.38 is discussed below.)
For de-reddening  we used a standard optical extinction curve \citep{card89}.  
%%%%%%%%%%%%%%%%%%%%%%% Fig 4 %%%%%%%%%%%%%%%%%%%%
\begin{figure}[t]
 \setlength{\unitlength}{1mm}
{\includegraphics[width=63mm, angle=-90, clip=]{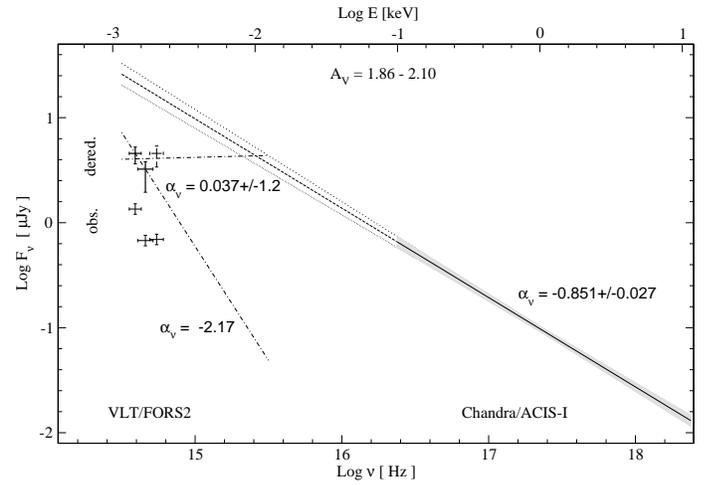}} 
 \caption{Tentative unabsorbed multi-wavelength spectrum for the inner part of the torus region of the G292.0+1.8 
 pulsar/PWN system compiled from the data obtained with the VLT and Chandra, as indicated in the plot. Thin cross-bars, 
 marked by "obs.", show the observed $VRI$ optical fluxes to demonstrate the interstellar extinction effect. 
 Error-bars for the de-reddened fluxes, marked as "dered", include $A_V$ uncertainty range shown at the top of the plot. 
 The $V$ band flux (dashed cross-bars) is likely contaminated by strong OIII emission from the SNR and may 
 be considered as a flux upper limit for the pulsar/PWN system. The grey polygon demonstrates the spectral uncertainty 
 in  X-rays.  It is extrapolated towards the optical  together with  the best X-ray power-law spectral fit  
 (solid and dashed lines). As seen, the extrapolated spectrum strongly overshoots the optical fluxes suggesting 
 at least one spectral break between the optical and X-rays. A crude power-law fits to the optical data 
 are shown by dot-dashed lines for two cases:when all three optical bands are included in the fit (upper line); 
 and when the $V$ band flux is excluded from the fit (low line with a higher slope). $\alpha_{\nu}$ are spectral indices of the power-law fits defined as  $F_{\nu} \propto
 \nu^{\alpha_{\nu}}$.   
  }
 \label{fig:mw}
 \end{figure}
%%%%%%%%%%%%%%%%%%% end Fig 4 %%%%%%%%%%%%%%%%%%%

%We combined the dereddened fluxes for the presumed optical counterpart of the pulsar/PWN system with 
%the unabsorbed X-ray spectrum in Fig.~\ref{fig:mw}. 
A tentative unabsorbed multi-wavelength spectrum of the central part of the pulsar/PWN system  
is compiled in Fig.~\ref{fig:mw}.
For this plot, we selected an $A_V$ range 
of 1.86--2.10 mag,  which covers the whole range of reasonable extinction values  %for the G292.0+1.8 center 
discussed above. The error-bars for the dereddened optical fluxes include this  % respective $A_V$ uncertainty 
range, while the centers of the cross-bars correspond to  de-reddenings with $A_V$$\approx$2.06 derived 
from  $N_H$ given by  
our X-ray spectral fit (cf. Table~\ref{t:phot}).  
{ We note, that the resulting dereddened flux errors are strongly dominated by the  $A_V$ uncertainties, and any  
 uncertainty  introduced by the star subtraction process are  insignificant at this level. }
 %  
% It is clear that the optical fluxes extrapolated from X-ray power-law spectrum overshoot strongly 
As seen, the extrapolation of the X-ray power-law spectral fit to the optical range 
overshoots  strongly  the optical fluxes. The coincidence can be achieved only at  unreasonable large extinction 
$A_V$$\ga$3.6 mag, which is significantly higher than the  entire  Galactic extinction in this direction, 
$A_V$$\approx$2.7 \citep{schleg98}. This suggests at least one spectral break between the optical and X-rays.
{ It is also  evident  for the pulsar spectrum if to combine the X-ray data and the  optical flux upper limit   obtained early by \citet{Hughes3}}. 

The actual number of the  breaks depends on the real spectral energy distribution (SED)  
in the optical range. As seen, both the observed and dereddened SEDs are apparently non-monotonic. 
One of the plausible reasons  is a contamination of the measured continuum emission  
form the pulsar/PWN in the $V$ band by { a strong background [OIII] emission from the oxygen-rich SN remnant 
knots and filaments \citep{wink06}. %(see Sect. 3.1). 
%strong  [OIII]  emission line from the remnant knots and filaments \citep{wink06} that enters into the $V$ passband.  
Some  emission clumps visible in this band south of the suggested counterpart may be also associated with faint [OIII] knots. 
Their presence   near the pulsar position cannot be excluded considering the available low spatial resolution  
narrow-band [OIII] images published by \citet{wink06}. Other emission lines (e.g., H$_{\alpha}$, [OI], [OII] and [SII]) detected 
from the SN remnant  in the red part of the spectrum enter into  the $R$ and $I$ passbands. However, they   
are much fainter than that of [OIII]  and cannot significantly contaminate the respective parts of the continuum  emission  
expected to be produced by the pulsar and its PWN. } 
%By these reasons, the  better coincidence of the shape  
%of the optical source in the red optical bands with the  X-ray PWN structure  is anticipated for 
%the real optical PWN counterpart. 
A similar situation has been observed in another young pulsar PSR B0540-69, 
which is also associated with an oxygen-rich SNR B0540-69.3 in the LMC \citep{ser04}. 
In this case, the flux in the $V$ band can be considered only as an upper limit for the 
pulsar/PWN emission. To underline this we represent %in the plot 
the respective dereddened 
flux by a dashed cross-bar in Fig.\ref{fig:mw}. 
If we exclude the $V$ band,
%Excluding this measurement,
   then the optical part of the spectrum becomes much steeper ($\alpha_{\nu}$$\approx$-2.2) 
than in X-rays ($\alpha_{\nu}$$\approx$-0.85), which suggests a double knee spectral break between these two ranges, 
as is observed  for the  PSR B0540-69/PWN system. Alternatively, if we take into account  the $V$ band as well, the resulting 
optical spectrum can be a flat or less steep, which leads to  a single spectral break, as is observed for the Crab PWN 
\citep[e.g.,][]{ser04,sol03}. %, although a less strong double knee break cannot be also excluded within large uncertainties  
%of the derived optical spectral index ($\alpha_{\nu}$$\approx$0.0$\pm$1.2).   

So far only the Crab and PSR B0540-69 PWNe have been firmly detected in the optical range. Within the uncertainties the shape of 
the tentative multi-wavelength spectrum of the %optical counterpart candidate to 
PSR J1124-5916/PWN system is either similar to  one  of the above systems or intermediate between them. 
This is an additional  
evidence of the correct optical identification. 
%%%%%%%%%%%%%%%%%%%%%%%     
%%%%%
%%%%%
%%%%%
%%%%%
\section{Discussion}
%%%%%%%%%%%%%%%%%%%%%%%%%%%%%%%%
\label{sec4}  
{ The deepest up to date broadband optical observations  of the PSR J1124-5916 field   
with the VLT  reveal a faint nebulosity at the pulsar position. 
The sub-arcsecond %remarkable 
%coincidence    
agreement
of  its brightness peak position  with the pulsar coordinates, 
 the likely similar optical source morphology  to that of  the brightest inner part of the  G292.0+1.8 X-ray torus-like PWN,  
 and   the consistence of its tentative multi-wavelength spectrum   
 with the spectra of other PWNe  allow us to consider the detected source as a very likely 
 %strongly  suggest that  it is   the 
 optical counterpart of the G292.0+1.8 pulsar$+$PWN  system. }
%%%%%%%%%%%%%%% Table  5  %%%%%%%%%%%%%%------------------------------
\begin{table*}[t] 
\caption{Comparison of the optical and X-ray spectral indices 
(${\rm \alpha^{O}_{\nu}}$, ${\alpha^{\rm X}_{\nu}}$) luminosities 
(${L^{\rm O}}$, ${L^{\rm X}}$), efficiencies (${\rm \eta ^{O}}$, 
${\rm \eta ^{X}}$) of the three young PWNe detected in the optical/IR 
and X-rays. Information on the Vela PWN, not yet detected in the optical, 
as well as on the pulsar characteristic ages ($\tau$), spindown pulsar 
luminosities (${\dot E}$), PWN sizes, and the ratios of the pulsar to PWN 
luminosity in the optical and X-rays are also included.    
}  
\label{t:pwn-lum}
\begin{center}
\begin{tabular}{lccccccccccc}
\hline\hline 
PWN & $\tau$ & ${\dot E}$ & size & ${\rm -\alpha^{O}_{\nu}}$ & ${L^{\rm O,a}}$ & ${\rm \eta ^{O}}$ & ${-\alpha^{\rm X}_{\nu}}$ 
& ${L^{\rm X,b}}$ & ${\rm \eta ^{X}}$ %& ${L^{\rm O}\Delta E^{\rm X} \over L^{\rm X}\Delta E^{\rm O}}$ 
& ${L^{\rm O}_{\rm psr}/L^{\rm O}_{\rm pwn}}$ & ${L^{\rm X}_{\rm psr}/L^{\rm X}_{\rm pwn}}$ \\  
  & kyr  & ${\rm 10^{37}~erg~s^{-1}} $  &pc & & ${\rm 10^{33}~erg~s^{-1}}$ & ${\rm 10^{-5}}$ &  & ${\rm 10^{36}~erg~s^{-1}}$ & ${\rm 10^{-3}}$ %&  
    &  & \\
\hline  
Crab$^c$  &1.24 &46     &1.5        & 0.92 & 4240   & 920        & 1.14  & 21.8                     & 47.5                       &0.0017    &0.046 
\widerul \\
0540$^c$  &1.66&50.2   &0.6--0.9   & 1.5  & 366    & 245        &  1.04  & 12                       & 79.7                       &0.03      &0.26 
\widerul  \\    
J1124$^d$ &2.9&1.2    &0.2--0.4 &0.0--2.2  & 0.02--0.09 & 0.17--0.76 & 0.85       & $4.2\times 10^{-3,f}$     & 0.34   
                   &$\la0.2$  & 0.21 
\widerul  \\ 
3C 58$^e$ &5.38&2.6    &0.08--0.19 &0.7--1.2  & 0.08--0.21 & 0.3--0.75 & 0.88       & $5\times 10^{-3}$            & 0.19                    &$\la0.1$  & 0.23 
\widerul  \\ 
Vela$^c$  &11&0.069  & 0.14      & --       & --         & --         & 0.5    & ${\rm 6.8\times 10^{-4}}$  & ${\rm 9.8\times 10^{-2}}$  & --       & 0.34
\widerul  \\
   
\hline
\end{tabular}
\label{t:pwn-params}
\begin{tabular}{ll} 
$^a$~For the optical range 1.57--3.68 eV. & $^d$~This paper. The distance is assumed to be 6.0 kpc. \\
$^b$~For the X-ray range 0.6--10 keV.      & $^e$~Data are taken from \citet{shib08}.   \\
$^c$~Data are taken from \citet{ser04}   & $^f$~This paper. For the whole torus region. \\                                                         \\
%$^c$~Optical fluxes are taken from Veron-Cety \& Woltjer (1993). &  range and $d=0.29$~kpc. \\
%$^d$~Kaaret et al. (2001); $ {L^{\rm X}}$ rescaled to the 0.6--10 keV range. & \\
\end{tabular} 
\end{center}
\end{table*} 
%%%%%%%%%%%%%%%  end Table 5  %%%%%%%%
{  The nebulosity appearance in the $VRI$ bands  excludes the possibility 
 that it is  due to  either a group of a few close unresolved faint field stars  or  
 overlapping wings  of the nearby detected stars.     

Alternatively it  could be a SNR filament, or  a faint distant background  galaxy 
 coinciding by chance with the pulsar position.      
%The correct identification is also supported by 
However,} 
 no  strong optical  and X-ray  emission lines typical 
for the SN remnant filaments were found within the central part of the PWN torus region \citep{wink06,park07}.  
This makes the filament interpretation rather unlikely. %shows that the detected optical source is not a remnant filament.   
%We also do not see  any signs of the lines  in the  residuals of 
%our absorbed power-law spectral fit to the X-ray emission from the center of this region. 
To
confirm this,
 %be confident in that, 
  we have analyzed  
%the X-ray ACIS-I data from 
the torus regions extended  up to $\sim$14\farcs8$\times$8\farcs6 
elliptical boundary using X-ray ACIS-I data.  A combination of the absorbed {\sl power-law} and {\sl vmekal} spectral  models 
and elliptical annular apertures centered on the pulsar was used. The contribution of the {\sl vmekal} 
becomes statistically significant only outside  $\approx$9\farcs6$\times$4\farcs8  
elliptical boundary, improving the spectral fit by the appearance of the O, Mg, 
and  Si emission lines from the soft thermal radiation of  a hot  SN remnant gas. This is roughly consistent 
with $\sim$5\asec$\times$3\asec~boundary, obtained by \citet{Hughes1} based on 
 previous ACIS-S data. Therefore,   the  strong line emission, typical of filaments, 
is at least  outside the limits of the optical source  and of the elliptical aperture used  to analyze 
the X-ray non-thermal emission from  the internal  torus region. 
On the other hand, further observations in the near-infrared and  near-UV  are necessary 
to better constrain the  SED of the nebulosity 
{  and to distinguish  it from 
a possible background galaxy.}  
The  power law  SED in a wide spectral range 
  would  be a strong confirmation  of the PWN origin. This would also enable us  
  to  better constrain  the extinction value towards the source and  to  check whether 
  the apparent flux excess in the $V$ band is indeed  due to the contamination  from  
  a faint and not yet resolved background [OIII] emission  of  the SN remnant. 

 Assuming that the pulsar contribution to the measured optical fluxes $\la$20\%,  based on the spatial profiles 
 of the nebulosity and on its surface brightness fits,  the predicted  optical magnitudes 
 of the point-like pulsar counterpart would be  $V$$\ga$26\fm2,  $R$$\ga$26\fm0, $I$$\ga$25\fm0.  
 These are quite reachable with the current telescope capacities. 
 However, higher spatial resolution is necessary to resolve  such  faint point-like object over the extended source  
 background
 {  and  also is important to minimize the contamination of the nebulous emission from the closest field 
 stars. % to better constrain its spatial structure  and measured  fluxes.  
 Deeper optical observations may 
 help to reveal outer fainter regions of the nebulosity at larger spatial  scale comparable 
 to the whole X-ray PWN extent.}        
 
There are no reports on radio detections from the central torus region of   J1124-5916 PWN. 
On  larger spatial scales, the G292.0+1.8 radio plerion  shows a core whose  brightness peak  
is located about 30\asec~north of the pulsar 
\citep{Gaen2}, while the pulsar is surrounded by  fainter  emission. 
It is interesting that the internal region of this radio core is  extended 
in the direction \citep[see Fig.~3 from ][]{Gaen2} roughly parallel to 
the  major axis of the elliptical PWN  torus region. At the same time, 
the outer contours of the radio core are extended  in the orthogonal direction, i.e.,  
south-south-west, which  may be associated with 
the direction of the X-ray jet propagating along the torus symmetry axis.  
This multi-wavelength structure of the J1124-5916 PWN appears to be  similar 
to what is seen for the Crab plerion that has a larger extent namely along 
the PWN torus symmetry axis. High resolution radio observations of G292.0+1.8 would 
be useful to  understand if the suggested similarity really exists.  
Such observations would also allow to constrain the radio flux from 
the compact PWN torus and its spectral slope in the radio range.

Assuming that we have  detected the optical counterpart of 
the torus-shaped PWN of G292.0+1.8, we compiled in  Table \ref{t:pwn-lum} 
the properties of this and other young PWNe detected in the optical and X-rays\footnote{Abbreviations: 0540 is B0540-69; J1124 is J1124-5916.}. 
We also include  the older Vela, whose PWN has not yet been identified 
in the optical range  \citep{mig03,shib03}. For G292.0+1.8 we adopt the distance 
of 6~kpc \citep{Gaen2}, and the  uncertainty  of its tentative PWN  optical luminosity 
is mainly due to the uncertainty in the interstellar extinction. We also updated 
the total X-ray luminosity of this PWN using the new  ACIS-I data and an elliptical 
aperture encapsulating the whole PWN torus flux. As a result, the updated flux is by an order 
of magnitude lower than it was estimated previously by \citet{Hughes1} using the ACIS-S data and 
assuming that the G292.0+1.8 PWN spectral index is the same as for the Crab PWN (see also \citet{kar08}). 
In the 2--10~keV image extracted from the  ACIS-I data, which is not contaminated 
by  radiation from strong SNR lines, one can resolve a faint diffuse non-thermal 
emission %extended 
around the PWN %up to the radius of 
extending
$\sim$30\asec--40\asec from the pulsar, 
partially overlapping with the radio core region. 
Accounting for this emission in the power law spectral fit with  $N_{H}$ fixed 
at the value derived for the central torus region,  
we obtained  an upper limit for the whole non-thermal X-ray luminosity 
of the plerion $L_{plerion}$$\approx$3$\times$10$^{34}$ erg~s$^{-1}$ (in 0.5-8~keV range). 
Compared with the whole torus luminosity,   we see that this faint but largely extended 
diffusive emission can contribute up to 85\%~of the total plerion luminosity. However, this %integral 
value  
is %anyway by 
a factor of two lower than %the 
early PWN luminosity estimates.       
The PSR J1124-5916 X-ray luminosity is taken from  \citet{Hughes2}, where 
the detection of the X-ray emission %pulsing 
with the pulsar period,  was reported. Our estimates show that it can be
%undefined 
uncertain 
by a factor of two. The respective value published later by \citet{kar08} actually coincides with 
the luminosity of the whole torus+pulsar system and, hence, is strongly overestimated. 
In Table \ref{t:pwn-lum} the pulsars are listed in characteristic age order. 
{  As  seen,   J1124-5916 fits well into the correlation %pulsar+PWN system 
of the X-ray and optical "strength" of PWNs with the spin-down luminosity. %, fading with age.
%nicely fits its position in this Table and does not disturb the overall evolution tendency when 
%the  X-ray and optical  "strength" of a PWN correlates with the spindown luminosity and fades with  age. 
 This  also supports %a supporting argument for 
 its correct optical identification.}
  The X-ray and tentative optical parameters of   J1124-5916 are very similar to the parameters of  3C~58 pulsar+PWN system, 
    recently identified in the optical and mid-infrared \citep{shib08,sla08,shear08}.
The coincidence of the properties of these two PWNe
%
%{  including those of the suggested optical counterpart to  J1124-5916},
% 
is not surprising, since their pulsars have similar key parameters like the spindown luminosity and age.  
By  chance, both torus-like PWNe are seen almost edge-on and display  one-side jets along the torus symmetry axis. 
Based that these two PWNe look like  twins,  
although the properties of the associated SNRs are very different.  
G292.0+1.8 is oxygen rich, while 3C~58 is not,  suggesting different masses of the SN progenitor stars.
%%%%%%%%%%%%%%%%%%%%%%%%%% Fig 5 %%%%%%%%%%%
\begin{figure}[]
\begin{center}
%\setlength{\unitlength}{1mm}
% \resizebox{7.cm}{!}{
% \begin{picture}(80,80)(0,0)
%\put (0,0) {
\includegraphics[width=84mm, bb=32 38 765 494,clip=]{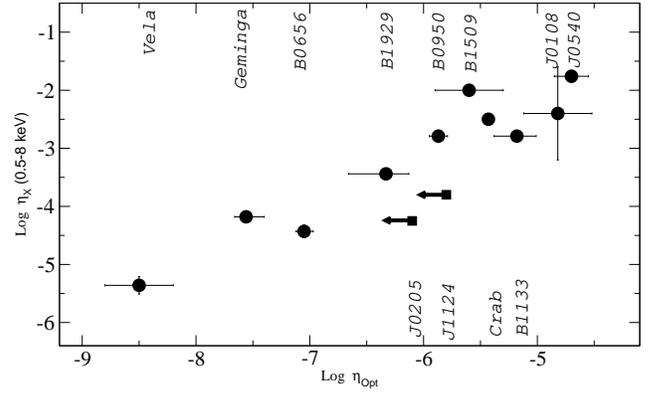}
%} 
% \end{picture}}
\end{center}
 \caption{ Relationship between the optical and X-ray efficiencies from \cite{zhar06} updated with data of PSR B1509-58 \citep{Wagner}, PSR B1133+14 \citep{zhar08}, PSR J0205+6449 \citep{shib08}, and PSR J0108-1431\citep{mig08}.  
 X-ray efficiencies (0.5-8keV) are based on non-thermal pulsar X-ray luminosities from \cite{kar08} and  Pavlov (private communication).}
 \label{fig:eff}
 \end{figure}
%%%%%%%%%%%%%% end Fig 5 %%%%%%%%%%%%%%

Finally,  we  check whether  the lower limit obtained here for the  optical luminosity of PSR J1124-5916 
and that of PSR J0205+6449 from \citet{shib08}
are compatible with the empirical relation between the non-thermal optical 
and X-ray efficiencies of the pulsars detected in both spectral domains \citep{zhar04,zhar06,zhar08}. 
This is shown in Fig.~\ref{fig:eff}.
% where we also included  the data for 
%PSR J0205+6449 in 3C~58  from  \citet{shib08}. 
As seen, the optical and X-ray data for  the two  pulsars 
do not contradict the relation. Both PSR J1124-5916 and PSR J0205+6449  naturally occupy 
an intermediate position in the plot  
between the younger Crab and B0540-69 pulsars and the older Vela, all of which   
power the torus-like PWNe. This ensure us that the  optical upper limits for both 
pulsars are realistic and close to the expected flux values based on their  X-ray  
and spindown luminosities. This encourages  further optical searching   
of both pulsars.           
 
In summary,   our  observations  of the PSR J1124-5916 field 
{  likely}   
increase 
the number of PWNe identified in the optical range from three to four. 
{  Additional evidences of the suggested  
identification  %of the J1124-5916 pulsar+PWN system  
should be provided 
by high spatial resolution optical observations. %are still necessary.
} 
 Together with the data in other spectral domains this should help for constraining models 
 and understanding the emission nature of these objects.   
 
%%%%%%%%%%%%%%%%%%%%
\begin{acknowledgements}
      We are grateful to anonymous referee for useful comments improving the paper. 
      The work was partially supported by CONACYT 48493 and PAPIIT IN101506 projects, RFBR (grants 08-02-00837a, 05-02-22003),
  NSh-2600.2008.2 and by by the German
      \emph{Deut\-sche For\-schungs\-ge\-mein\-schaft, DFG\/} project
      number Ts~17/2--1. We used the USNOFS Image and Catalogue Archive operated by the United States Naval Observatory, Flagstaff Station (http://www.nofs.navy.mil/data/fchpix/).  The Munich Image Data Analysis System is developed and maintained by the European Southern Observatory.      
IRAF is distributed by the National Optical Astronomy Observatories,
    which are operated by the Association of Universities for Research
    in Astronomy, Inc., under cooperative agreement with the National
    Science Foundation.
\end{acknowledgements}
%%%%%%%%%%%%%%%%%%%%%%%%%%%%5
%%%%%%%%%%%%%%%%%


\begin{thebibliography}{}
%%%%%%%%%%%%%%%%%%%%%

\bibitem [Camilo et al.(2002)] {Camilo} Camilo, F., Manchester, R. N., Gaensler, B. M., et al. 2002, 
\apj, 567, L71

\bibitem [Cardelli et al.(1989)]{card89}  Cardelli, J. A., Clayton, G. C., 
Mathis, J. S. 1989, \apj, 345, 245 

\bibitem [Chevalier(2005)]{chev05}   Chevalier, R. 2005,  \apj, 619, 839 

\bibitem[Gaensler et al.(2002)]{Gaen2} Gaensler, B. M., Arons, J., Kaspi, V.M., Pivovaroff, M.J., Kawai, N. 2002, \apj, 569, 878

\bibitem[Gaensler \& Wallace(2003)]{GW}Gaensler, B.M., \& Wallace, B. J. 2003, \apj, 594, 326

\bibitem [Hughes et al.(2001)] {Hughes1}  Hughes, J. P., Slane, P. O., Burrows, D. N., 
Garmire, G. 2001, \apj, L153

\bibitem [Hughes et al.(2003)] {Hughes2} Hughes, J. P., Slane, P. O., Park, S., et al. 2003, 
\apj, 591, L139 

\bibitem [Hughes et al.(2004)] {Hughes3}  Hughes, J. P.,  Friedman, R. B., Slane, P. O., Park, S.  2004,% IAU Symposium 218, 199
Young Neutron Stars and Their Environments, IAU Symposium no. 218, held as part of the IAU General Assembly, 14-17 July, 
2003 in Sydney, Australia. Edited by Fernando Camilo and Bryan M. Gaensler. San Francisco, CA: Astronomical Society of the Pacific, 2004., 199

\bibitem [Hwang, Holt \& Petre(2000)]{hwan00} Hwang, U., Holt, S., \&  Petre, R. 2000, \apj 537, L119 

\bibitem[Fukugita et al.(1995)]{Fukugita}Fukugita, M., Shimasaku, K., \& Ichikawa, T. 1995, \pasp, 107, 945

\bibitem[Kargaltsev \& Pavlov(2008)]{kar08} Kargaltsev, O., \& Pavlov, G.G. 2008, astro-ph 0801.26002

\bibitem[Landolt(1992)]{Landolt}Landolt, A.  1992, \aj, 104, 340  

\bibitem[Mignani et al.(2003)]{mig03} Mignani, R.P.,  De Luca, A.,  Kargaltsev, O. et al. 2003, \apj, 594, 419
 
\bibitem[Mignani(2005)]{mig05} Mignani, R.P.  2005, Proc. of "The Electromagnetic Spectrum of Neutron Stars", 
  eds. A. Baykal et al., Springer 210, 133 

\bibitem[Mignani et al.(2008)]{mig08}Mignani, R.P., Pavlov, G.G, \& Kargaltsev, O. 2008, arXiv:0805.2586

\bibitem[Park et al.(2007)]{park07} Park, S., Hughes, J.P., Slane, P.O., Burrows, D.N., Gaensler, B.M., Ghavamian P. 2007, 
  \apj, 670, L121 

\bibitem[Predehl \& Schmitt(1995)]{pred95} Predehl, P. \& Schmitt, J.H.M.M. 1995, A\&A, 293, 889


\bibitem[Safi-Harb \& Gonzalez(2002)]{safi02} Safi-Harb, S. \& Gonzalez, M.E. 2002, in:  "X-rays at Sharp Focus": Chandra Science Symp. 
 ASP Comf. Series Vol 262, eds. E. Schlegel and S.D. Vrtilek  

\bibitem [Schlegel et al.(1998)]{schleg98} Schlegel, D. J., Finkbeiner, D. P., Davis, M.  1998, \apj, 500, 525 

\bibitem [Serafimovich et al.(2004)]{ser04}    Serafimovich, N., Shibanov, Yu.A., Lundqvist, P., Sollerman, J. 
 2004,   A\&A, 425, 1041 

\bibitem[Shearer  \&  Neustroev(2008)]{shear08} Shearer, A. \&  Neustroev, V.V.  
2008,  MNRAS, accepted  (astro-ph:0807.0983v1)
 
\bibitem[Shibanov et al.(2003)]{shib03} Shibanov, Yu.A., Koprsevich, A.B, Sollerman, J., Lundqvist, P. 
2003,  A\&A, 406, 645	


\bibitem[Shibanov et al.(2008)]{shib08} Shibanov, Yu.A., Lundqvist, N., Lundqvist, P., Sollerman, J., 
Zyuzin, D. 2008,  A\&A,  486, 273 
%astro-ph 0802.2386

\bibitem [Slane et al.(2008)]{sla08}    Slane, P., Helfand, D. J., Reynolds. S.P., et al. 2008, \apj, 676, L33


\bibitem  [Sollerman(2003)]{sol03}   Sollerman, J.  2003,  A\&A, 406, 639 


%\bibitem [Serafimovich et al.(2004)]{ser04}    Serafimovich, N., Shibanov, Yu.A., Lundqvist, P., Sollerman, J. 
% 2004,   A\&A, 425, 1041

\bibitem[Zavlin \& Pavlov(2004)]{zav04}Zavlin, V. E., Pavlov, G. G. 2004, \apj, 616, 452

\bibitem[Zharikov et al.(2002)]{zhar02} Zharikov, S., Shibanov, Yu.,  Koptsevich, A.  et al. 2002,  A\&A, 394, 633

\bibitem[Zharikov et al.(2004)]{zhar04} Zharikov, S.,  Shibanov, Yu., Mennickent R.,  et al. 2004, A\&A, 417, 1017

\bibitem[Zharikov et al.(2006)]{zhar06} Zharikov, S., Shibanov, Yu., Komarova, V. 2006, AdSpR, 37, 1979

\bibitem[Zharikov et al.(2008)]{zhar08} Zharikov, S., Shibanov, Yu., Mennickent R., \& Komarova, V. 2008, A\&A, 479,
793 

\bibitem[Wagner \& Seifert(2000)]{Wagner} Wagner, S. J., \& Seifert, W. 2000, Pulsar Astronomy - 2000 and Beyond, 
ASP Conference Series,  202,  Proceedings of the 177th Colloquium of the IAU held in Bonn, 
Germany, 30 August - 3 September 1999.  
Edited by M. Kramer, N. Wex, and N. Wielebinski, 315

\bibitem[Winkler \& Long(2006)]{wink06} Winkler, P.F., Long, K.S. 2006, \aj, 132, 360   

\end{thebibliography}
\end{document}